\documentclass[prr,twocolumn]{revtex4}
\usepackage{amsmath,amssymb}
\usepackage{amsthm}
\usepackage{epsfig}
\usepackage{xcolor}
\usepackage{fixltx2e}
\usepackage[normalem]{ulem}
\usepackage{siunitx}

\usepackage{xr}
\graphicspath{{eps/}}            % folder with figures

\newcommand{\RFRe}[1]{\mathrm{RF}^{\mathrm{Re}}_{#1}}
\newcommand{\RFIm}[1]{\mathrm{RF}^{\mathrm{Im}}_{#1}}

\newcommand{\RFnRe}[1]{\mathrm{RF}^{\mathrm{Re}}_{n#1}}
\newcommand{\RFnIm}[1]{\mathrm{RF}^{\mathrm{Im}}_{n#1}}

\newcommand{\KnRe}[1]{K^{\mathrm{Re}}_{n#1}}
\newcommand{\KnIm}[1]{K^{\mathrm{Im}}_{n#1}}

\newcommand{\iu}{\mathrm{i}}

\begin{document}
% version for arXiv
\title{Eigenvalue-Based Approach to Manipulate and Reconstruct Nonlinear Pulses: Towards Soliton Tomography}

\author{Sergey Dremov$^{1}$}
\author{Rustam Mullyadzhanov$^{1,2,3}$}
\author{Andrey Gelash$^{1}$}
\email[Corresponding author : ]{a.gelash@skoltech.ru}

\affiliation{$^{1}$Center for Engineering Physics, Skolkovo Institute of Science and Technology, Moscow, Russia}
\affiliation{$^{2}$Novosibirsk State University, Novosibirsk, Russia}
\affiliation{$^{3}$Institute of Thermophysics SB RAS, Novosibirsk, Russia}

\begin{abstract}
Soliton content of nonlinear pulses of different physical nature is universally characterized by a discrete set of eigenvalues. In an ideal channel governed by the nonlinear Schrodinger equation, the eigenvalues do not change along the wave field propagation. Perturbations leave predictable fingerprints on the eigenvalue portrait, which was recently used to manipulate optical fiber solitons in [Phys. Rev. Lett. 134, 193804, 2025]. Here, we develop a theoretical framework to manipulate and reconstruct sech-shaped nonlinear wave fields based on soliton eigenvalue response functions and the corresponding inverse problem. We derive analytical expressions to enable nonlinear manipulation of solitons by applying instant, controllable perturbations. Then we present a concept of perturbation sensing with the key feature of nonlinear propagation of the probe signal over an unknown distance, enabling the extraction of information about the perturbation source hidden within nonlinear media or materials. We introduce an integral equation for the inverse problem of reconstructing the unknown shape of the wave field distortions, when the known observational data is a function of deviations in soliton eigenvalues measured at the end of the nonlinear propagation channel. We evaluate different reconstruction regimes and demonstrate a reliable inverse problem solution in presence of noise, paving the way towards soliton tomography.
\end{abstract}
\maketitle
\section{Introduction}
The integrable nature of the nonlinear Schrödinger equation (NLSE) model underlies many nonlinear wave phenomena in various physical systems, such as elastic collisions of solitons, propagation of breathing wave complexes and formation of extreme amplitude rogue waves with rational profiles \cite{zakharov1972exact,NovikovBook1984,AblowitzBook1981,copie2025controlled,xu2019breather,kharif2008rogue,akhmediev2009rogue}. From practical point of view, the integrability has found application, e.g., in the concept of telecommunications based on the encoding of information into the eigenvalues of solitons, which are preserved or nearly preserved during the long-haul propagations of optical signals \cite{yousefi2014information,turitsyn2017nonlinear,le2017nonlinear}. Theoretical description of the NLSE relies on the Inverse Scattering Transform (IST) theory and the concept of scattering data \cite{zakharov1972exact,NovikovBook1984}, the latter playing the role of a nonlinear analogue of the Fourier spectra of wave fields \cite{AblowitzBook1981}. The IST theory provides exact formulas for single and multi-soliton dynamics, as well as an asymptotic behavior of nonlinear evolution for an arbitrary-shaped pulses. Extended with a perturbation technique, the IST covers a broad spectrum of soliton physics in optics, hydrodynamics, plasma, and Bose-Einstein condensates \cite{KarpmanJETP1977,KaupSIAM1976,KivsharRMP1989}. When a purely analytical description proves ineffective or becomes too complex, the numerical version of IST, also called the nonlinear Fourier transform (NFT), helps to analyze the composition of scattering data, and synthesize initial wave profiles with desired nonlinear properties \cite{OsborneBook2010,wahls2015fast,gelash2018strongly,suret2020nonlinear,fache2025dissipation}.

Here we advance the approach, recently proposed in \cite{mucci2025manipulation}, of the IST control of nonlinear pulses and unlock the potential of soliton eigenvalues in nonlinear sensing of signal distortions and nonlinear media perturbations. Our theoretical model is the focusing one-dimensional NLSE:
\begin{eqnarray}\label{eqNLS}
	i q_t + \frac12 q_{xx} + |q|^2 q = 0,
\end{eqnarray} 
where $q(x,t)$ is complex wave field envelope function, $t$ is evolution variable (time) and $x$ is spatial coordinate. The IST method for the NLSE is rooted in auxiliary Zakharov-Shabat (ZS) eigenvalue problem \cite{zakharov1972exact} written on the two-component wave function $\mathbf{\Phi} = (\phi_1, \phi_2)^T$ at a fixed moment of time
\begin{equation}\label{ZSh-eigenvalue}
	\widehat{\mathcal{L}}\mathbf{\Phi} = \lambda \mathbf{\Phi},
	\quad
	\widehat{\mathcal{L}} = i \begin{pmatrix}\ 1 & 0 \\ 0 & -1 \end{pmatrix}\frac{\partial}{\partial x} - i\begin{pmatrix}\ 0 & q \\ q^* & 0 \end{pmatrix},
\end{equation}
where $\lambda = \xi + \iu\eta$ is a complex-valued spectral parameter and $\widehat{\mathcal{L}}$ is the so-called Lax operator. Discrete eigenvalues $\{\lambda_n = (-V_n + i A_n)/2\}$ of the system (\ref{ZSh-eigenvalue}) correspond to solitons having amplitudes $A_n$ and velocities $V_n$ embedded into the wave field $q(x)$, which plays the role of potential in the ZS system. The eigenvalues are conserved quantities of the NLSE model and can be changed by an instant wave field perturbation or nonintegrable corrections to the model (\ref{eqNLS}).

Previously we introduced soliton eigenvalue response functions $\text{RF}_n(x)$ for box-shaped wave fields, featuring specific combination of the wave function $\mathbf{\Phi}_n$ components, and applied these formalism in optical fiber experiments to manipulate strongly interacting solitons embedded into box-shaped light field \cite{mucci2025manipulation}. Similar to the probability amplitude in quantum mechanics \cite{landau1958quantum}, the response functions (RFs) reveal how perturbations instantaneously applied to a nonlinear pulse affect specific solitons, even though the latter may be completely delocalized due to mutual overlapping. The RFs reveal the spatial sensitivity of the multi-soliton fields, allowing one to manage individual soliton velocities and amplitudes and even selectively separate optical solitons from a propagating light pulse \cite{mucci2025manipulation}.

Here in the first part of the work, we develop soliton eigenvalue response functions formalism for sech-shaped nonlinear pulses, which is an analytically solvable case close to optical fiber and other physical experiments. Using exact solutions to the ZS problem obtained in the work \cite{satsuma1974b}, we derive sech-potential RFs and demonstrate soliton manipulation in numerical modeling of integrable nonlinear pulse propagation based on Eq.~(\ref{eqNLS}). In the second part of the work we consider a fundamental question: can the information contained in the deviations of soliton eigenvalues enables to reconstruct the applied perturbation when nonlinear modifications of the pulse do not allow this to be done directly? To answer the question, we formulate the inverse eigenvalue response problem, expressed in terms of integral equations with kernel $K_{n}(x,A)$ constructed from the RFs at various amplitudes $A$ of the probe pulse. In this approach, by interrogating pulses of different known amplitudes, we probe a fixed external perturbation and reconstruct it using methods naturally employed in inverse problems, paving the way towards soliton tomography in different physical systems: from fiber optics to water surface waves and plasmas.

\section{Theoretical background}
Without loss of generality, we fix the moment of time $t=0$. The auxiliary ZS eigenvalue problem (\ref{ZSh-eigenvalue}) can be rewritten in the form of scattering problem for potential $q=q(0,x)$ as
\begin{equation}\label{the_ZS_problem}
    \mathbf{\Phi}_x = \begin{pmatrix}
        -\iu\lambda & q \\
        -q^* &  \iu\lambda
    \end{pmatrix} \mathbf{\Phi}
\end{equation}
with the boundary conditions imposed on the wave function $\mathbf{\Phi} = (\phi_1, \phi_2)^T$
\begin{equation}\label{BD}
        \begin{pmatrix}
        \phi_1 \\
        \phi_2
    \end{pmatrix} \xrightarrow[x \to -\infty]{} \begin{pmatrix}
         e^{\iu\lambda x}\\
        0
    \end{pmatrix}, ~ \begin{pmatrix}
        \phi_1 \\
        \phi_2
    \end{pmatrix} \xrightarrow[x \to \infty]{} \begin{pmatrix}
         a(\lambda)e^{\iu\lambda x}\\
        b(\lambda)e^{-\iu\lambda x} 
    \end{pmatrix},
\end{equation}
where $a(\lambda)$ and $b(\lambda)$ are scattering coefficients.

In our work, we consider as initial conditions sech-shaped wave fields
\begin{equation}\label{A_sech}
    q(0,x) = A \operatorname{sech}(x)
\end{equation}
with real-valued amplitude $A$. As was shown in \cite{satsuma1974b}, system (\ref{the_ZS_problem}) with potential (\ref{A_sech})
can be solved analytically. We write the corresponding solution of the wave function, satisfying boundary conditions (\ref{BD}) in the following form
\begin{align}
    &\phi_1(s) = s^{\iu\lambda/2} (1-s)^{-\iu\lambda/2}\times \nonumber\\
    &{}_2F_1 (-A,A,1/2 - \iu\lambda, 1-s), \nonumber\\
    &\phi_2(s) = \frac{s^{(1 +\iu\lambda)/2}(1-s)^{(1-\iu\lambda)/2}}{A^{-1}(\iu\lambda - 1/2)} \times \nonumber\\
    &{}_2F_1 (1-A,1 + A,3/2 - \iu\lambda, 1-s).
    \label{Eigf}
\end{align}
Here $s = (1-\tanh(x))/2$, and ${}_2F_1(\alpha,\beta,\gamma,z)$ is a hyper-geometric function. Note, that for the purposes of further derivations we represent solution (\ref{Eigf}) in a modified form compared to the one obtained in \cite{satsuma1974b}.

The scattering coefficients, which follow from solution (\ref{Eigf}) and boundary conditions (\ref{BD}), are
\begin{eqnarray}\label{ab_coeffs}
    a(\lambda) &=& \frac{\Gamma^2(1/2 - \iu\lambda)}{\Gamma(1/2 - \iu\lambda + A)\Gamma(1/2 - \iu\lambda - A)},
    \nonumber\\
    b(\lambda) &=& \frac{-\sin(\pi A)}{\cosh(\pi\lambda)},
\end{eqnarray}
where $\Gamma(z)$ is the gamma function. The set of discrete (soliton) eigenvalues $\{\lambda_n\}$ of the sech-potential is defined by the zeros of the scattering coefficient $a(\lambda)$:
\begin{equation}\label{Eig_DS}
    \lambda_n = \iu(A - n - 1/2),  \quad n = 0,\,1, \dots, \lfloor A - 1/2 \rfloor,
\end{equation}
where $\lfloor\cdot\rfloor$ denotes the floor function. At discrete eigenvalue points $\lambda_n$, where $a(\lambda_n)=0$, the corresponding wave functions of the ZS system $\mathbf{\Phi}_n(x) = (\phi_{n,1}, \phi_{n,2})^T$ are bounded as $|x| \rightarrow \infty$ due to cancellation of the growing exponent in the right asymptotic of Eq.~(\ref{BD}).

As follows from Eq.~(\ref{Eig_DS}) the eigenvalues are purely imaginary; therefore, wave functions (\ref{Eigf}) are purely real for solitons. Note that when increasing $A$, a new $n$-th soliton emerges at $A = n+1/2$, and between these consecutive birth points, the number of solitons embedded in the pulse does not change. In addition to discrete eigenvalues, each soliton is characterized by its so-called norming constant $\rho_n$:
\begin{equation}\label{rho_n}
    \rho_n = \frac{b(\lambda)}{a'(\lambda)} \bigg|_{\lambda = \lambda_n} = \frac{-i~\Gamma(2A - n)}{n!~\Gamma(A - n)},
\end{equation}
which completes discrete scattering data set. The explicit connection of the scattering data with solitons' amplitude $A_n$, velocity $V_n$, position $x^{n}_0$, and phase $\theta^{n}_0$ is as follows:
\begin{eqnarray}\label{connect_Eigv}
    A_n &=& 2\eta_n,
    \nonumber\\
    V_n &=& -2\xi_n,
    \nonumber\\
    \rho_n &=& -\iu A_ne^{A_nx^{0}_n - i \phi^{0}_n}.
\end{eqnarray}

According to the well established IST perturbation theory \cite{KarpmanJETP1977,KaupSIAM1976}, an instant perturbation $\delta q(x)$ of initial potential $q(x)$ leads to change in eigenvalues $\delta \lambda$ described as
\begin{equation}
\delta\lambda = \frac{\langle \Phi^{\dagger}, \delta\hat{L} \Phi \rangle}{\langle \Phi^{\dagger}, \Phi \rangle}, \qquad 
\delta\hat{L} = -\iu \begin{pmatrix}
0 & \delta q \\
\delta q^* & 0
\end{pmatrix}.
\end{equation}
As follows from Eq.~(\ref{connect_Eigv}), varying eigenvalues $\lambda_n$ alters the amplitude and velocity of the $n$-th soliton, thereby enabling the manipulation of solitons within nonlinear wave pulses.

\section{Eigenvalue response functions for solitons}

Following \cite{mucci2025manipulation}, we introduce soliton eigenvalue response functions (RFs) based on the bounded wave functions of the ZS system obtained at discrete eigenvalue points $\lambda_n$:
\begin{align}\label{RFS}
    \RFnRe{}(\lambda_n,x) &= \frac{-\iu}{\Delta(\lambda_n,x) } \Big(\phi^2_1(\lambda_n,x) + \phi^2_2(\lambda_n,x) \Big), \cr
    \RFnIm{}(\lambda_n,x) &= \frac{1}{\Delta(\lambda_n,x) } \Big(\phi^2_2(\lambda_n,x) - \phi^2_1(\lambda_n,x) \Big), \cr
    \Delta(\lambda_n,x) &= 2\int^{\infty}_{-\infty} \phi_1(\lambda_n,x) \phi_2(\lambda_n,x)~dx.
\end{align}

The eigenvalue RFs facilitate a more convenient application of the perturbation theory. For example, if an instant potential perturbation is $\delta q = \delta q^{\mathrm{Re}} + \iu \delta q^{\mathrm{Im}}$, then the changes of the $n$th soliton eigenvalues $\delta \lambda_n$ are given by the following overlap integrals
\begin{equation}\label{Eigs_RF}
    \delta \lambda_n = \int^{\infty}_{-\infty} \RFnRe{} \delta q^{\mathrm{Re}} ~dx + \int^{\infty}_{-\infty} \RFnIm{} \delta q^{\mathrm{Im}} ~dx.
\end{equation}

Using (\ref{Eigf}) in (\ref{RFS}), we obtain general expressions for RFs of the sech-shaped potential. In the case of discrete spectrum eigenvalues, hyper-geometric functions ${}_2F_1$ in expressions (\ref{Eigf}) reduce to ${}_2F_1 (-A,A,A - n, 1-s)$ and ${}_2F_1 (1-A,1+A,A - n + 1, 1-s)$, which depend on $A$ and $n$ only, enabling an analytical formulation of soliton manipulation approach. In what follows, we derive exact expressions for $n=0$ and $n=1$, where explicit analysis is feasible.

\subsection{The fundamental soliton}
First, we consider the fundamental soliton characterized by $n = 0$, $\lambda_0 = \iu(A - 1/2)$, and $A \geq 1/2$. Substituting these discrete spectrum data into (\ref{Eigf}) and applying the property ${}_2F_1 (a,b,b,z) = (1-z)^{-a}$ yields
\begin{align}\label{Eigf_0}
    \phi^2_1(s) &= s^{A + 1/2} (1-s)^{A - 1/2}, \cr
    \phi^2_2(s) &= s^{A - 1/2} (1-s)^{A + 1/2}, \cr
    \phi^2_2 \pm\phi^2_1 &= [s(1-s)]^{A - 1/2} \cdot (1-s \pm s).
\end{align}
In terms of variable $s$, function $\Delta$ at $\lambda=\lambda_0$ becomes
\begin{equation}
    \Delta(s) = -\int^1_0 [s(1-s)]^{A-1} ds = -\mathrm{B}(A,A),
\end{equation}
where $\mathrm{B}(x,y)$ is the beta function. Using $\tanh(x) = 1 - 2s$ and $4s(1-s) = \operatorname{sech}^2(x)$ finally gives the RFs of the fundamental soliton
\begin{align}\label{RF_DS_0}
        \RFRe{n=0}(x) &= \frac{\iu}{\mathrm{B}(A,A)} \left(\frac{\operatorname{sech}(x)}{2}\right)^{2A-1}, \cr
        \RFIm{n=0}(x) &= \frac{-1}{\mathrm{B}(A,A)} \left(\frac{\operatorname{sech}(x)}{2}\right)^{2A-1} \tanh(x).
\end{align}

We highlight that, as follows from (\ref{RF_DS_0}), $\RFRe{n=0}(x)$ is \textit{imaginary} and \textit{even}, characterizing the sensitivity of the soliton to amplitude changes, whereas $\RFIm{n=0}(x)$ is \textit{real} and \textit{odd}, characterizing its sensitivity to velocity changes. This is consistent with the general properties of the soliton perturbation theory.

\subsection{The second soliton}

For later use, we also consider the second soliton with $n = 1$, $\lambda_1 = \iu(A - 3/2)$, and $A \geq 3/2$. Following a similar procedure as we used for the fundamental soliton and omitting the intermediate steps, we present the expressions for the RFs of the second soliton in the sech-shaped pulse:
\begin{align}\label{RF_DS_1}
    \RFRe{n=1} &= \RFRe{n=0}\cdot\Bigl[\frac{4(A - 1)^2\sinh^2(x) +1}{2A-1}\Bigr], \cr
    \RFIm{n=1} &= \RFIm{n=0}\cdot\Bigl[\frac{4(A-1)^2\cosh^2(x)}{2A-1} - 2A + 1 \Bigr]. \cr
\end{align}
RFs (\ref{RF_DS_1}) also retain the required properties of being real- or complex-valued, and having parity.

Considering subsequent solitons $(n > 1)$ yields cumbersome expressions; therefore, it is more practical to calculate them numerically. Note, that qualitatively, the RFs of solitons from the sech potential resemble those for the box potential case \cite{mucci2025manipulation}. For example, the $\mathrm{RF}^{\mathrm{Im}}_n(x)$ are zero at the center and have $n$ zeros on the left and right, separated by local minima and maxima, indicating regions of maximum and minimum soliton sensitivity.

\subsection{Numerical computation of eigenvalue RFs}
For most potential shapes, analytical solutions to the ZS problem are inaccessible; instead, the wave functions can be computed numerically using one of the available direct scattering transform approaches \cite{boffetta1992computation,wahls2015fast,mullyadzhanov2019direct,medvedev2024fast}. To illustrate this opportunity, we numerically recompute the RFs for the potential $q = 2.1\operatorname{sech}(x)$ shown in Fig.~1. We use the second-order Boffetta-Osborne method \cite{boffetta1992computation} to solve problem (\ref{the_ZS_problem}) at fixed discrete eigenvalue points $\lambda_0 = 1.6\iu$ and $\lambda_1 = 0.6\iu$. Then we construct RFs of the fundamental and second solitons from Eqs.~(\ref{RFS}). When choosing computational parameters, one should account for subtleties in the numerical solution of the scattering problem caused by the presence of exponentially large and small values in the direct scattering transform scheme \cite{Gelash2020}. We choose a spatial domain of length $L = 40$, discretized with $N = 512$ points, which is sufficient to enable an accurate comparison between analytical and numerical results. We demonstrate the coincidence of theoretical and numerical RFs in Fig.\ref{RFs_comparison} by adding numerical points over analytical curves.

\subsection{Manipulation of solitons using eigenvalue RFs}
\begin{figure}[t!]
\center{\includegraphics[width=0.5\textwidth]{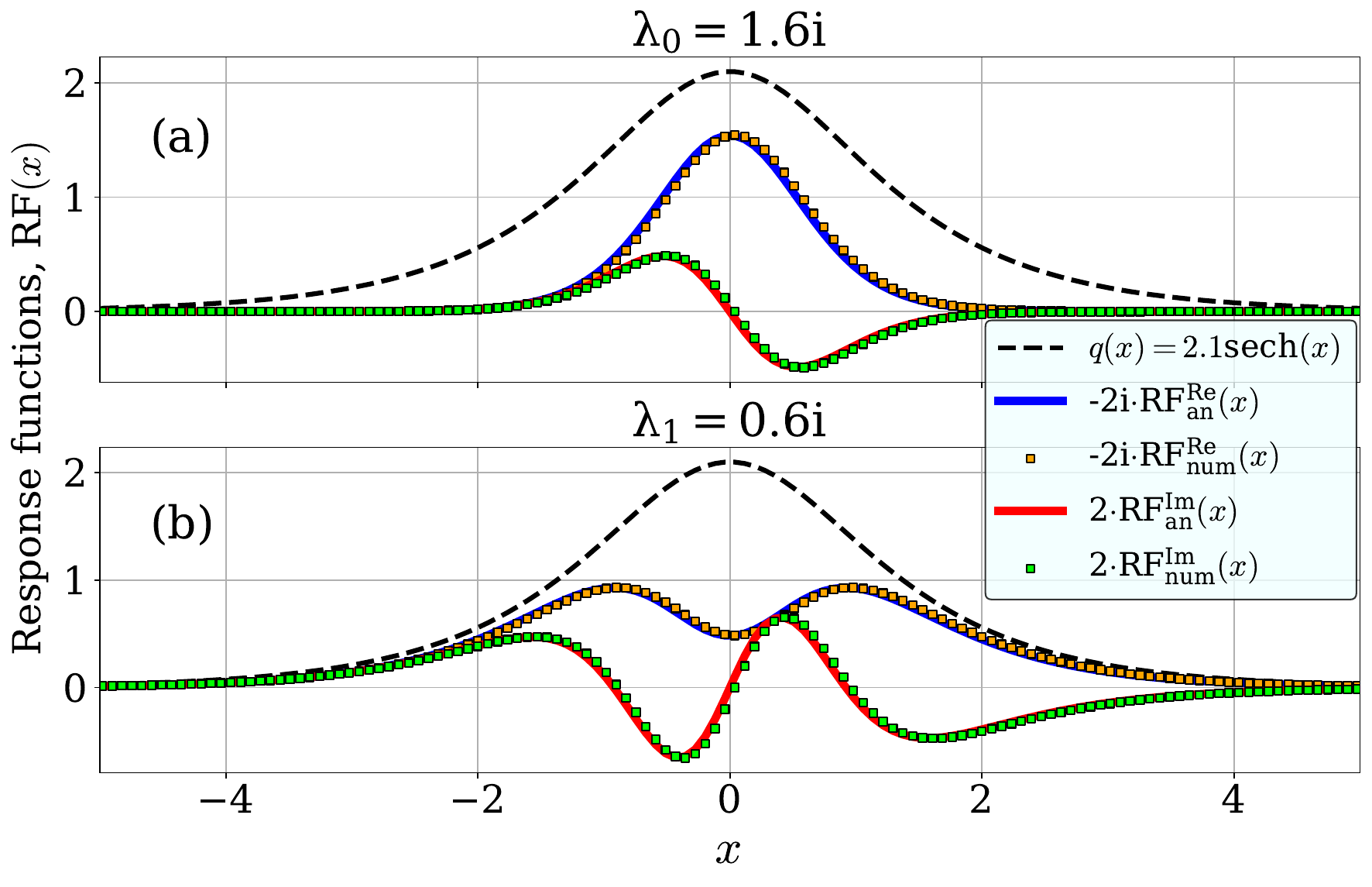}}
\caption{Soliton eigenvalue RFs for potential $q = 2.1\operatorname{sech}(x)$ (dashed line) obtained analytically via (\ref{RF_DS_0}) and (\ref{RF_DS_1}) (solid blue and red lines), and numerically (orange and green markers). Panels (a) and (b) show the comparison for the two discrete eigenvalues of the pulse: $\lambda_0 = 1.6\iu$ and $\lambda_1 = 0.6\iu$, respectively.}
\label{RFs_comparison}
\end{figure}

Now we demonstrate how to apply our RFs framework using a couple of benchmark potentials $q = 2.1\operatorname{sech}(x)$ and $q = 3.3\operatorname{sech}(x)$. The first 
wave field contains two solitons with eigenvalues $\lambda_0 = 1.6\iu$ and $\lambda_1 = 0.6\iu$, and a portion of continuous spectrum. The corresponding complete set of soliton eigenvalue RFs, computed using Eqs.~(\ref{RF_DS_0})~and~(\ref{RF_DS_1}), are shown in Fig.\ref{RFs_comparison}. 

\begin{figure}[h!]
\center{\includegraphics[width=0.48\textwidth]{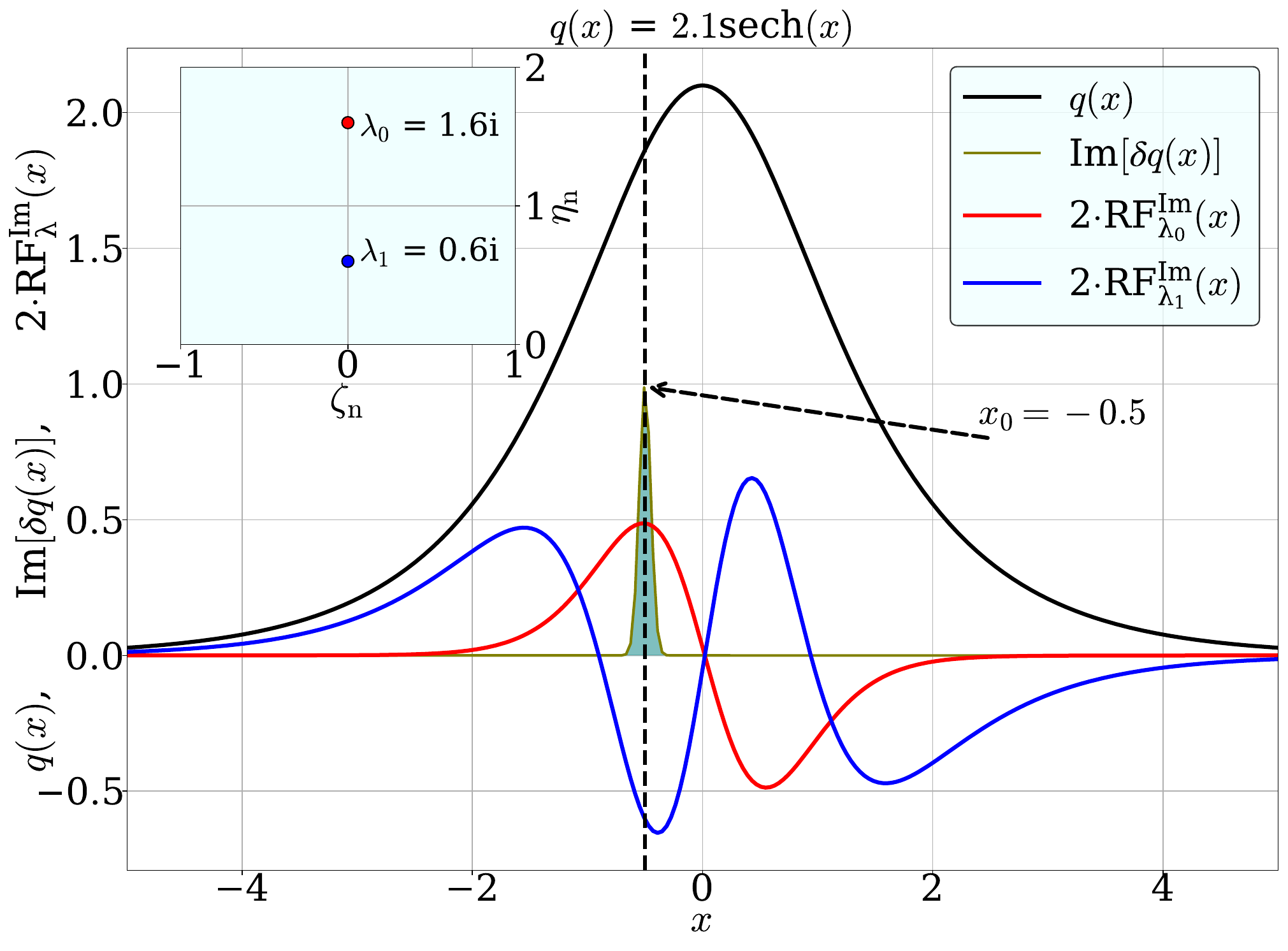}}
%\center{\small (a)}
\caption{Unperturbed potential $q(x) = 2.1\operatorname{sech}(x)$ (solid black line) containing two solitons with $\lambda_0 = 1.6\iu$ and $\lambda_1 = 0.6\iu$. Response functions $2 \cdot \RFIm{}$ are plotted for $\lambda_0$ (red) and $\lambda_1$ (blue). The Gaussian perturbation $\delta q$ (olive with teal shading) with $a = 1.0$ and $\sigma = 0.05$ is applied at $x_0 = -0.5$ (dashed vertical line and arrow). At this point, the RFs are near their extrema and have opposite signs, suggesting a subsequent near-symmetric decay of the initial state to two separate solitons with velocities $-2\delta \lambda_0$ and $-2\delta \lambda_1$, respectively.}
\label{2.1sech_RF}
\end{figure}

\begin{figure}[h!]
\center{\includegraphics[width=0.5\textwidth]{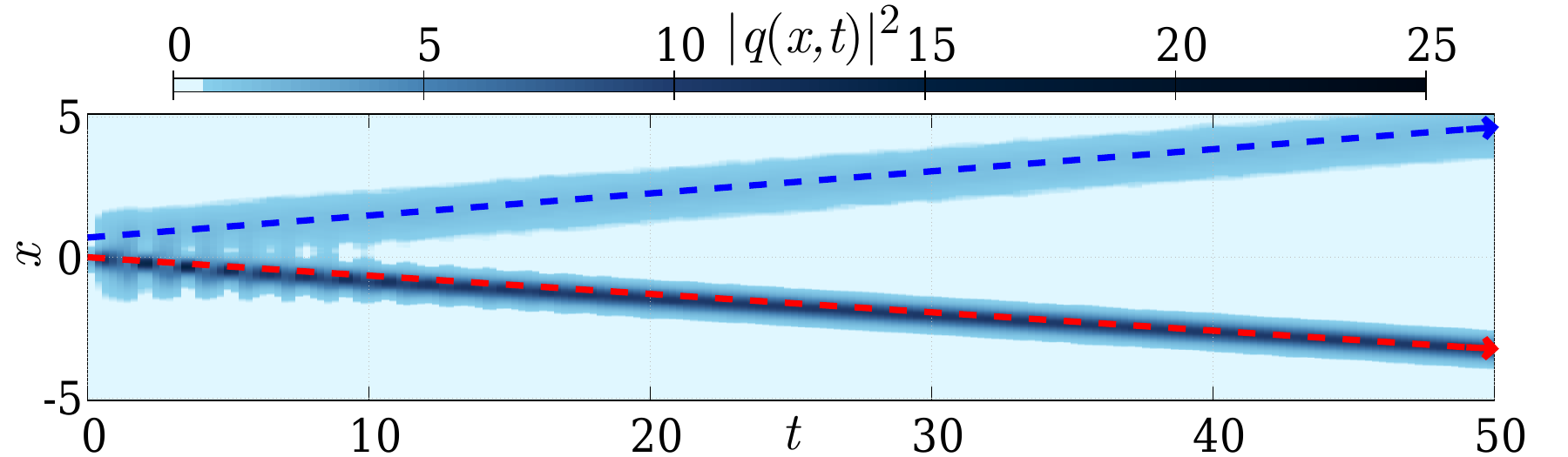}}
%\center{\small (b)}
\caption{The $(x\text{-}t)$-diagram of the SSFM simulation for potential $q(x) = 2.1\operatorname{sech}(x)$ perturbed by the Gaussian pulse $\delta q$ ($a = 1.0, \sigma = 0.05, x_0 = -0.5$) under the NLSE. The predicted in Fig.\ref {2.1sech_RF} soliton decay can be clearly observed. Red and blue dashed lines indicate velocities predicted using the RFs for solitons with $\lambda_0$ and $\lambda_1$, respectively.}
\label{2.1sech_SSFM}
\end{figure}

Consider the same potential $q = 2.1\operatorname{sech}(x)$ with two solitons inside, we apply imaginary perturbation $\delta q = \iu\cdot\delta q^{\mathrm{Im}}(x)$, where $\delta q^{\mathrm{Im}}(x)$ is a Gaussian pulse with arbitrary amplitude $a$, width $\sigma$, and position $x_0$. According to (\ref{Eigs_RF}) in the first order of the perturbation theory, an imaginary perturbation changes only the velocities of the solitons. While $a$ and $\sigma$ in $\delta q$ play the role of a scaling factor, varying $x_0$ results in different scenarios for the subsequent dynamics of the perturbed wave field. A proper choice of $x_0$ can ensure that the initial pulse separates into individual solitons moving in opposite directions. We verify these predicted scenarios by simulating the evolution of perturbed initial conditions under the NLSE model using the Split-Step Fourier Method (SSFM). The theoretical predictions can be tested directly from an $(x\text{-}t)$-diagram by comparing trajectories of the solitons with theoretical expressions for their velocities. An example of initial conditions, including the potential, its soliton RFs, and the perturbation with $a = 1.0, \sigma = 0.05$ applied at $x_0 = -0.5$, is shown in Fig.~\ref{2.1sech_RF}, while the corresponding SSFM evolution, which confirms the predictions, is presented in Fig.~\ref{2.1sech_SSFM}.

Note that the eigenvalue RFs formalism captures only the slope of the soliton trajectories, while their origin, i.e., positions of solitons after the perturbation applied, is defined by the change of soliton norming constants $\rho_n$; see Eqs.~(\ref{rho_n}) and (\ref{connect_Eigv}). The IST perturbation theory allows for computing these changes $\delta\rho_n$ by, for example, solving additional differential equations \cite{mullyadzhanov2021solitons}. Meanwhile, the possibility of representing $\delta\rho_n$ in the convenient form of overleaping integrals remains an open question.

\begin{figure}[h!]
\center{\includegraphics[width=0.48\textwidth]{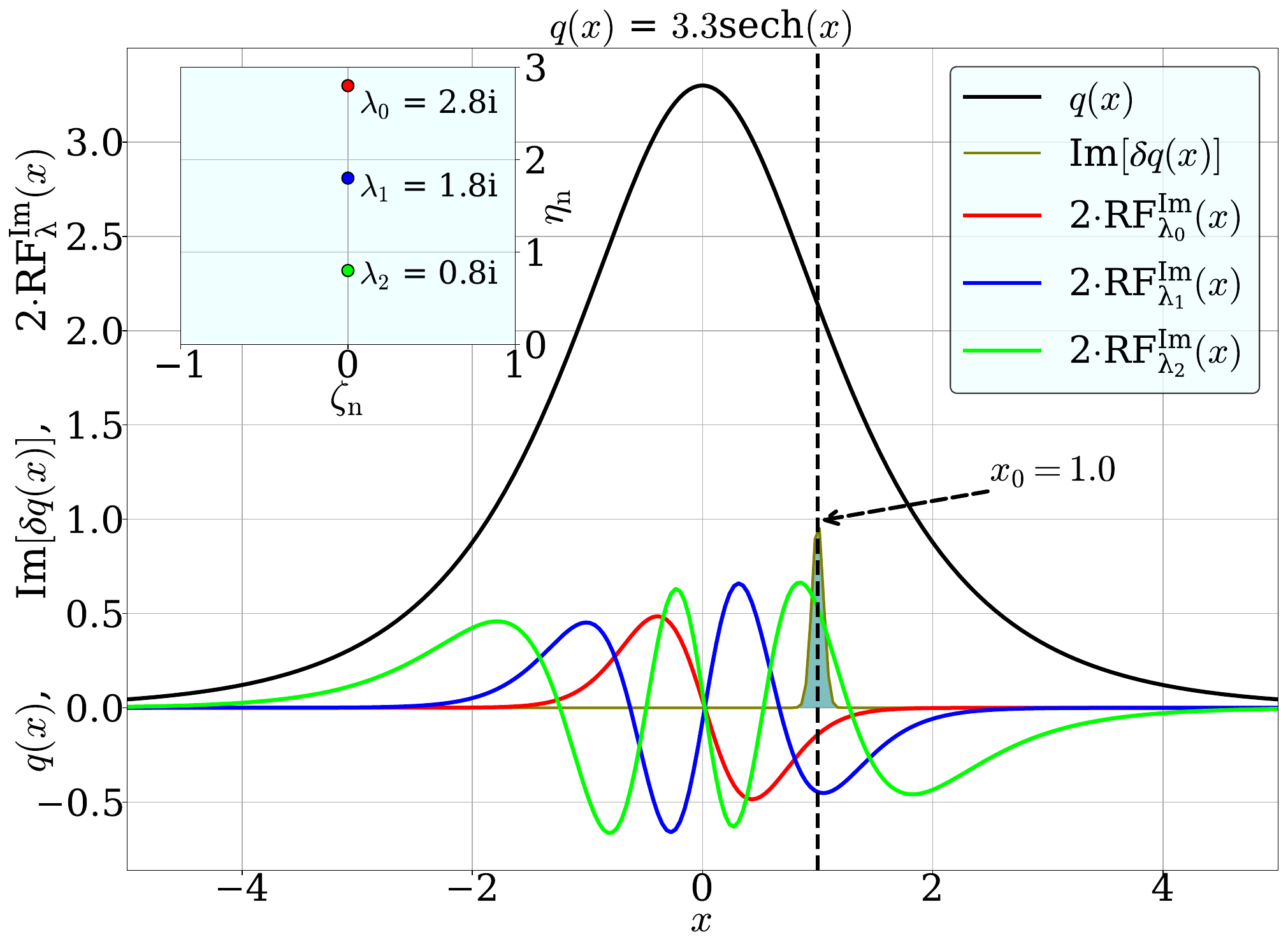}}
%\center{\small (a)}
\caption{Unperturbed potential $q(x) = 3.3\operatorname{sech}(x)$ (solid black line) containing three solitons with $\lambda_0 = 2.8\iu$, $\lambda_1 = 1.8\iu$, $\lambda_2 = 0.8\iu$. Response functions $2 \cdot \RFIm{}$ are plotted for $\lambda_0$ (red), $\lambda_1$ (blue), and $\lambda_2$ (green). The Gaussian perturbation $\delta q$ (olive with teal shading) with $a = 1.0$ and $\sigma = 0.05$ is applied at $x_0 = 1.0$ (dashed vertical line and arrow). At this point, the RFs approach predicts a decay of the initial wave field to three separate solitons propagating with different velocities $-2\delta \lambda_0$, $-2\delta \lambda_1$, and $-2\delta \lambda_2$, respectively.}
\label{3.3sech_RF}
\end{figure}

\begin{figure}[h!]
\center{\includegraphics[width=0.5\textwidth]{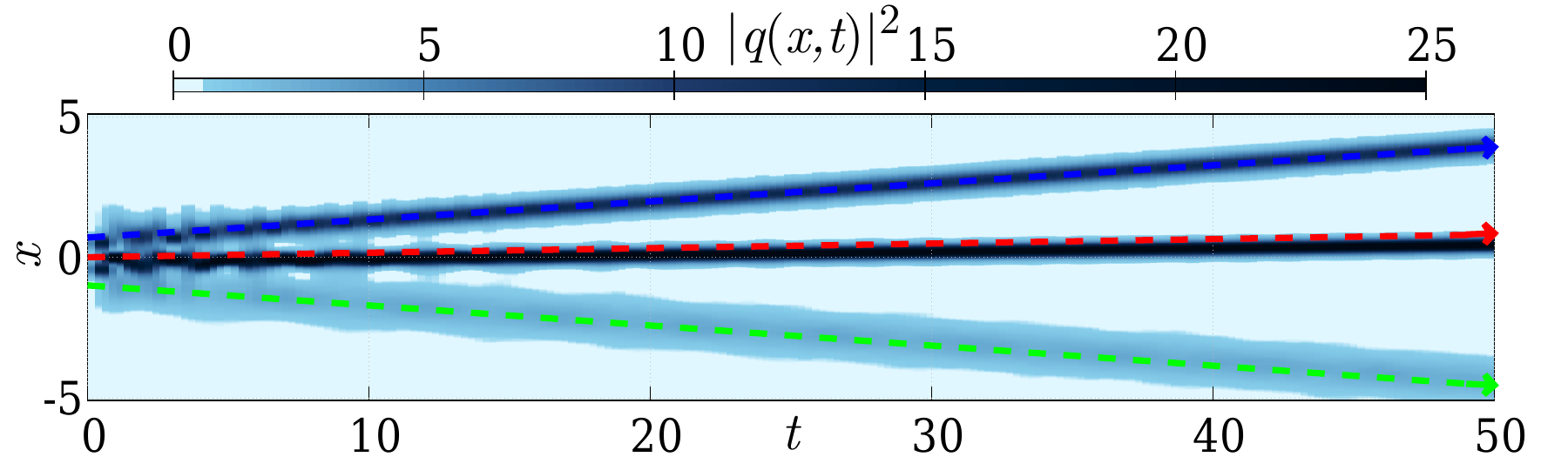}}
%\center{\small (b)}
\caption{The $(x\text{-}t)$-diagram of the SSFM simulation for potential $q(x) = 3.3\operatorname{sech}(x)$ perturbed by the Gaussian pulse $\delta q$ ($a = 1.0, \sigma = 0.05, x_0 = 1.0$) under the NLSE. The predicted in Fig.\ref{3.3sech_RF} soliton decay can be clearly observed. Red, blue and green dashed lines indicate velocities predicted using the RFs for solitons with $\lambda_0$, $\lambda_1$, and $\lambda_2$, respectively.}
\label{3.3sech_SSFM}
\end{figure}

Finally, as a more complex example, we consider potential $q(x) = 3.3 \operatorname{sech}(x)$ with three solitons characterized by eigenvalues $\lambda_0 = 2.8\iu $, $\lambda_1 = 1.8\iu$, and $\lambda_2 = 0.8\iu$. The potential also contains a continuous spectrum. The soliton structure of this potential, together with the same perturbation as considered before, $\delta q$ ($a = 1.0, \sigma = 0.05$) applied now at the point $x_0 = 1.0$, is presented in Fig. \ref{3.3sech_RF}. In this case, RFs analysis predicts a decay to three solitons with different velocities $-2~\delta\lambda_n$. The corresponding SSFM evolution of the perturbed initial wave field, which again verifies our theory, is shown in Fig \ref{3.3sech_SSFM}. 

Compared with previously considered approaches to splitting multi-soliton pulses into individual solitons \cite{satsuma1974b,tai1988fission,kaup1994effect,prilepsky2007breakup,gao2026soliton}, the RF formalism yields overlapping integrals (\ref{Eigs_RF}), enabling semi-analytical analysis of the eigenvalue-manipulation task and facilitating convenient soliton manipulation in complex nonlinear wave fields.

\section{Inverse Problem for the Pulse Perturbation Reconstruction}
Representing eigenvalue deviations as overlap integrals (\ref{Eigs_RF}) opens the door to formulating the inverse problem for reconstructing the shape of perturbations by interrogating the medium with a series of probe pulses having different soliton eigenvalue portraits. Here, in the second part of our work, we formulate the inverse problem by introducing integral equations whose kernel $K(x,y)$ is constructed from the eigenvalue RFs, where $y$ is a pulse varying parameter. Then we present a concept of perturbation sensing with the key feature of nonlinear propagation of the probe signal over an unknown distance, enabling the extraction of information about the perturbation source hidden within nonlinear media or materials.

\subsection{Formulation of the inverse problem}
We choose amplitude $A$ in Eq.~(\ref{A_sech}) as a varying parameter to generate set of probe pulses and rewrite integral equations 
based on overlap integrals (\ref{Eigs_RF}) as
\begin{eqnarray}\label{Eigs_RF_A}
    &&\delta \lambda_n (A) = \int^{\infty}_{-\infty} \KnRe{}(x,A)\cdot\delta q^{\mathrm{Re}}(x) ~dx +
    \\\nonumber
    &&\int^{\infty}_{-\infty} \KnIm{}(x,A)\cdot\delta q^{\mathrm{Im}}(x) ~dx,
\end{eqnarray}
where $K^{\mathrm{Re/Im}}_{n}(x,A) \equiv \mathrm{RF}^{\mathrm{Re/Im}}_{n}(x,A)$ is our notation for the kernel of the integral equation constructed from the response functions at varying pulse amplitude $A$.

Now suppose we have a set of data $\boldsymbol{\delta \lambda}_n = \{\delta \lambda_n(A_i)\}$ for different probe pulses $A_i\operatorname{sech}(x)$, where $A_i$ takes values $\{A_1, \dots, A_M\}$, and $M$ is a total number of the corresponding datasets, each obtained at fixed $A_i$. The pulse perturbation $\boldsymbol{\delta q} = \{ \delta q (x_j) \}$ remains identical for different $A_i$ and is now unknown. Here, $x_j \in \{x_1, \dots, x_N\}$ represents the grid points discretizing domain $L$ into $N$ points. Then, the discrete formulation of integral equations (\ref{Eigs_RF_A}) yields
\begin{equation}\label{Fredholm_Eq}
    \boldsymbol{\delta \lambda}_n = \boldsymbol{K}^\mathrm{Re}_{n} \cdot \boldsymbol{\delta q}^{\mathrm{Re}} + \boldsymbol{K}^\mathrm{Im}_{n} \cdot \boldsymbol{\delta q}^{\mathrm{Im}},
\end{equation}
where $\boldsymbol{\delta \lambda}_n$ is a $M$-column vector, $\boldsymbol{\delta q}^{\mathrm{Re/Im}}$ are $N$-column vectors, and $\boldsymbol{K}^\mathrm{Re/Im}_n = \{\mathrm{RF^{Re/Im}_n}(A_i, x_j)\}$ are $M \times N$ matrices.
Eq.~(\ref{Eigs_RF_A}) and its discretized version, Eq.~(\ref{Fredholm_Eq}), are Fredholm integral equations of the first kind with kernels based on the response functions $\RFRe{}(A,x)$ and $\RFIm{}(A,x)$. The inverse problem is ill-posed: the perturbation $\boldsymbol{\delta q}$ to be found is sensitive to small deviations in $\boldsymbol{\delta \lambda}$, making it impossible to find a unique solution without additional assumptions or employing regularization techniques \cite{doicu2010numerical}. In a general situation, the number of data sets $M$ obtained at different probe signal amplitudes is different from the number of signal discretization points $N$. When $M < N$, system (\ref{Fredholm_Eq}) is underdetermined, while when $M > N$ it is overdetermined. 

\begin{figure*}[!t]\centering
	\includegraphics[width=17.75cm]{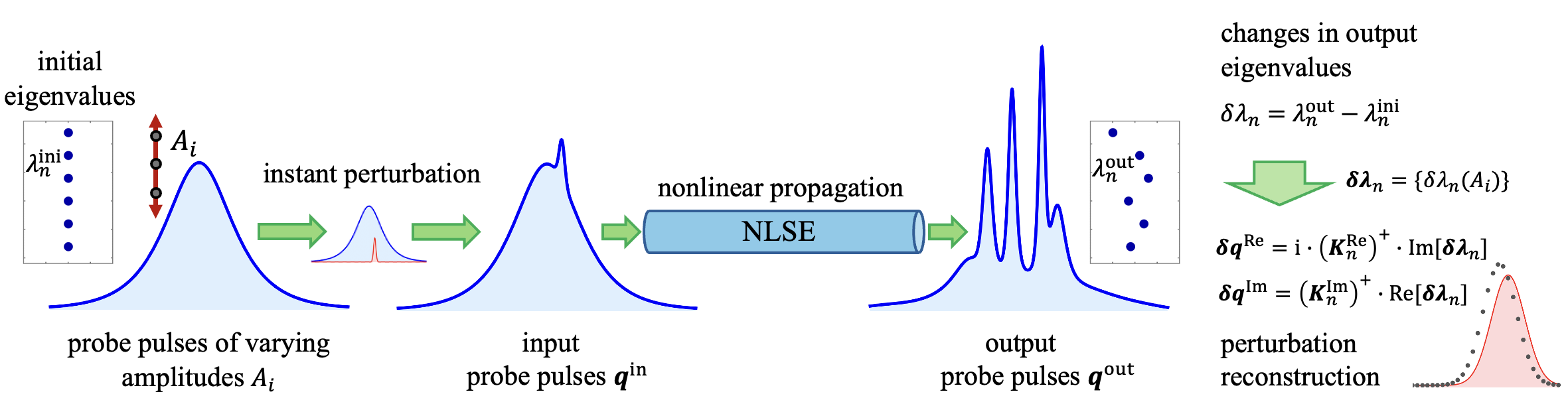}
	\caption{Illustration for the concept of perturbation reconstruction via the inverse problem for changes in soliton eigenvalues. A series of probe nonlinear multi-soliton pulses undergoes the same perturbation instantaneously, leading to systematic changes in the soliton eigenvalues. After nonlinear propagation over an (in general) unknown distance, the eigenvalues are detected and used as input data to solve the inverse problem of the perturbation reconstruction.}
\label{fig:inv-concept}
\end{figure*}

Note that the inverse problem based on symmetric probe pulses, such as $q(x)=q(-x)$, has indistinguishable solutions symmetric with respect to each other, making the reconstruction of an arbitrary perturbation $\delta q(x)$ inaccessible. Mathematically, this means that since $\Phi(A,x) = \Phi(A,-x)$, then the same symmetry applies to response functions: $\mathrm{RF^{Re/Im}}(A,x) = \mathrm{RF^{Re/Im}}(A,-x)$. The latter results in a linear dependency of the column vectors in $\boldsymbol{K}^\mathrm{Re/Im}_n$, and hence in a zero determinant of these matrices. To overcome these restrictions, one needs to use asymmetric probe potentials. Here, as a proof of concept, we consider only sech potential (\ref{A_sech}) and to make the inverse problem solvable we restrict it to the semi-infinite interval $x \geq 0$ and decompose $\boldsymbol{\delta q} = \boldsymbol{\delta q^+} + \boldsymbol{\delta q^-}$ with $\boldsymbol{\delta q^+}$ defined on $x \geq 0$, and $\boldsymbol{\delta q^-}$ on $x < 0$, and subsequently set $\boldsymbol{\delta q^-} = 0$. In other words, we assume that the perturbation is applied to the right part of the probe pulse only.  In a general case, to lift the symmetry restrictions, one can use an asymmetric double-box potential for which an analytic solution to the ZS problem is available or an arbitrary asymmetric wave field with numerically computed soliton eigenvalue response functions.

As we saw in the first part of the work, perturbations leave distinctive and well-understood fingerprints in soliton eigenvalues, making them a promising source of information for the inverse problem solving. Using properties of the eigenvalue response functions discussed in Sec.~III we derive from Eq.~(\ref{Fredholm_Eq}) a system of two equations for $\mathrm{Re}[\boldsymbol{\delta q}]$ and $\mathrm{Im}[\boldsymbol{\delta q}]$:
\begin{align}\label{dq_split_system}
    \mathrm{Im}[\boldsymbol{\delta \lambda}_n] &= -\iu \cdot \boldsymbol{K}^\mathrm{Re}_{n} \cdot \boldsymbol{\delta q}^{\mathrm{Re}}, \cr
    \mathrm{Re}[\boldsymbol{\delta \lambda}_n] &= \boldsymbol{K}^\mathrm{Im}_{n} \cdot \boldsymbol{\delta q}^{\mathrm{Im}}.
\end{align}
Now, reconstructing $\boldsymbol{\delta q} = \boldsymbol{\delta q}^{\mathrm{Re}} + i \boldsymbol{\delta q}^{\mathrm{Im}}$ boils down to inverting matrices $\boldsymbol{K}^\mathrm{Re}_{n}$ and $\boldsymbol{K}^\mathrm{Im}_{n}$ in (\ref{dq_split_system}). In general case $M \neq N$ the inverse matrix is not defined; therefore, we use the Moore–Penrose pseudo-inverse formalism implemented via Singular Value Decomposition (SVD), see e.g. \cite{axler2024linear}:
\begin{equation}\label{SVD}
    \boldsymbol{K} = \boldsymbol{U} \boldsymbol{\Sigma} \boldsymbol{V}^\dagger,
\end{equation}
where $\boldsymbol{K}$ stands for $\boldsymbol{K}^\mathrm{Re}_{n}$ or $\boldsymbol{K}^\mathrm{Im}_{n}$. $\boldsymbol{U}$ and $\boldsymbol{V}^\dagger$ are complex unitary matrices of size $M \times M$ and $N \times N$, respectively, with the dagger denoting Hermitian conjugation. $\Sigma$ is a rectangular diagonal $M \times N$ matrix such that $\Sigma_{ii} = \sigma_i$ for $i = 1, \dots, \mathrm{min}(M,N)$, and all other elements are zero. Values $\sigma_1 \ge \sigma_2 \ge \dots \ge \sigma_{\mathrm{min}(M,N)} \ge 0$ are called singular values of matrix $\boldsymbol{K}$. Then pseudo-inverse matrix $\boldsymbol{K}^+$ can be found as
\begin{equation}\label{SVD_inv}
    \boldsymbol{K}^+ =\boldsymbol{V} \boldsymbol{\Sigma}^+ \boldsymbol{U}^\dagger,
\end{equation}
where $\boldsymbol{\Sigma}^+$ is a $N \times M$ matrix obtained by transposing $\boldsymbol{\Sigma}$ and replacing each non-zero singular value $\sigma_i$ with its reciprocal $1/\sigma_i$. Now using (\ref{dq_split_system}) we obtain the final expressions to reconstruct the probe pulse perturbation:
\begin{align}\label{dq_reconstruct}
    \boldsymbol{\delta q}^{\mathrm{Re}} &= \iu \cdot (\boldsymbol{K}_n^\mathrm{Re})^{+} \cdot \mathrm{Im}[\boldsymbol{\delta \lambda}_n], \cr
    \boldsymbol{\delta q}^{\mathrm{Im}} &= (\boldsymbol{K}_n^\mathrm{Im})^{+} \cdot \mathrm{Re}[\boldsymbol{\delta \lambda}_n].
\end{align}

Soliton eigenvalue RFs (\ref{RF_DS_0}) and (\ref{RF_DS_1}) enable explicit access to soliton integral kernels $\boldsymbol{K}^\mathrm{Re}_{n}$ and $\boldsymbol{K}^\mathrm{Im}_{n}$, simplifying subsequent analysis of the inverse problem. As follows from Eq.~(\ref{dq_split_system}) each $n$th soliton of the probe pulse provides an independent kernel, i.e., more information about the perturbation vector $\boldsymbol{\delta q}$. In this work, we limit ourselves to the fundamental and first soliton kernels with $n=0$ and $n=1$, and below compare the efficiency of the inverse perturbation reconstruction for each.

As follows from the behavior of RFs (\ref{RF_DS_0}) and (\ref{RF_DS_1}), soliton kernels $\boldsymbol{K}$ are smooth, and therefore their singular value spectrum decays exponentially. The corresponding matrices contain very small singular numbers $\sigma_i$; upon inversion, these produce small denominators in (\ref{SVD_inv}) and large numerical errors, directly manifesting the ill-posed nature of the original problem. This feature cannot be eliminated entirely; however, the standard way to mitigate its consequences is to employ a matrix kernel regularization procedure. From a physical perspective, this means imposing certain constraints on the perturbation itself and incorporating apriori knowledge of its properties, such as smoothness, the absence of discontinuities, and finiteness. The choice of the regularization scheme and its parameters constitutes a separate, highly non-trivial, and vast topic, see e.g. \cite{doicu2010numerical}. Here we use one of the best-known method: ridge regression, or Tikhonov regularization \cite{tikhonov1963solution,tikhonov1977solutions}. The method smooths out all small singular values $\sigma_i$ of $\boldsymbol{K}$, since they typically do not correspond to any physical solutions. Calculating $\boldsymbol{\Sigma}^+$ in (\ref{SVD_inv}) is now performed by replacing $1/\sigma_i$ with $\sigma_i/(\sigma^2_i + \alpha^2)$, where $\alpha > 0$ is a regularization parameter, that yields
\begin{equation}\label{Kernel_reg}
    \boldsymbol{K}^+_{\mathrm{reg}} = \boldsymbol{V} \boldsymbol{\Sigma}^+_{\mathrm{reg}} \boldsymbol{U}^\dagger,
\end{equation}
which for $\sigma_i \gg \alpha$ results in no regularization ($\sigma_i/(\sigma^2_i + \alpha^2) \approx 1/\sigma_i$), while for $\sigma_i \ll \alpha$ the expression behaves like $\sigma_i/\alpha^2$ suppressing the unstable components.

\subsection{Pulse perturbation reconstruction}

According to the IST theory, once perturbed, soliton eigenvalues retain their values during signal propagation along a nonlinear channel governed by the NLSE, allowing us to formulate a concept of perturbation reconstruction via the inverse problem for changes in soliton eigenvalues; see Fig.~\ref{fig:inv-concept}. We assume that our nonlinear system consists of a pulse generator, followed by an instant perturbation source, and a long-distance propagation channel. For pulses detected after propagation, one must numerically solve the ZS eigenvalue problem using one of the available algorithms and identify soliton eigenvalue changes. An experimental realization of such a system is possible, for example, using a recirculating optical fiber loop setup with phase modulators, as reported in \cite{mucci2025manipulation}, or in other nonlinear systems that enable control of the injected pulse shapes.

\begin{figure}[t!]
    \centering
    \includegraphics[width=1.01\linewidth]{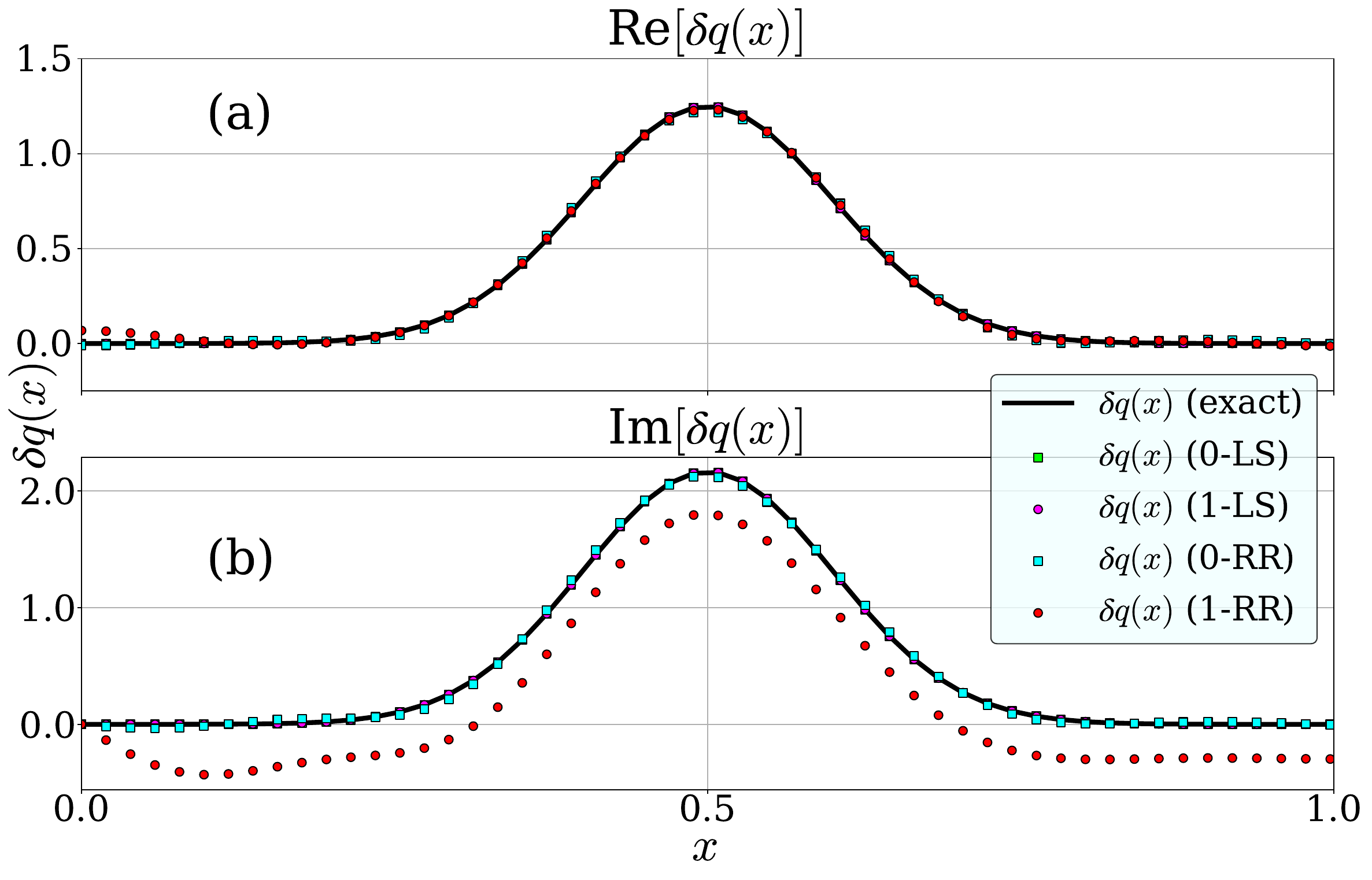}
    \caption{Reconstructing a single Gaussian perturbation $\boldsymbol{\delta q}$ ($a = 2.5, \sigma = 0.1, x_0 = 0.5$, and $\theta = \pi/3$, exact) from theoretical data $\boldsymbol{\delta\lambda}$. Discrete points show inverse problem solution using the the least squares (LS) method, and the ridge regression (RR) based on the first ($n = 0$) and the second ($n = 1$) soliton kernels. Panels (a) and (b) present the real and imaginary parts of the Gaussian perturbation, respectively.}
    \label{Single_Gaussian_Ex_Rec}
\end{figure}

As a proof of concept, we consider a single Gaussian as a benchmark perturbation. In contrast to the first part of the work, now the perturbation is more general: it is complex-valued with an arbitrary phase $\theta$:
\begin{equation}\label{dq_1gauss}
    \delta q(x) = ae^{-(x - x_0)^2/2\sigma^2} \cdot e^{i \theta}.
\end{equation}
We specify a set of $M = 16$ perturbed input probe pulses $q^{\text{in}}(x) = A_i \operatorname{sech}(x) + \delta q(x)$ with $A_i \in [3, 10]$ equally spaced. The lower bound $A_1 = 3$ is chosen to allow the use of the second soliton RF (\ref{RF_DS_1}) in addition to the fundamental one. The spatial domain is defined for $x \in [0,L]$ with $L=20$ and discretized into $N = 1024$ points. First, we solve an idealized inverse problem assuming the vector of theoretical eigenvalue deviations $\boldsymbol{\delta\lambda}$ to be exactly known. We perform the inverse reconstruction using ridge regression (RR) with $\alpha = 10^{-13}$, with the SVD implemented in the SciPy library \cite{virtanen2020scipy}. Additionally, we use the least squares (LS) method to reconstruct $\delta q$ by calculating optimal $a, \sigma, x_0, \theta$ for (\ref{dq_1gauss}) to minimize the residual in (\ref{Fredholm_Eq}). The LS method is a simple, versatile approach to solving the inverse problem, which is available to us when the perturbation $\delta q(x)$ can be parametrized by a few parameters. An example of reconstructing $\boldsymbol{\delta q}$ with $a = 2.5, \sigma = 0.1, x_0 = 0.5$, and $\theta = \pi/3$ from exact $\boldsymbol{\delta\lambda}^{\text{ex}}$ is shown in Fig. \ref{Single_Gaussian_Ex_Rec}. 

\begin{figure}[h!]
    \centering
    \includegraphics[width=1.01\linewidth]{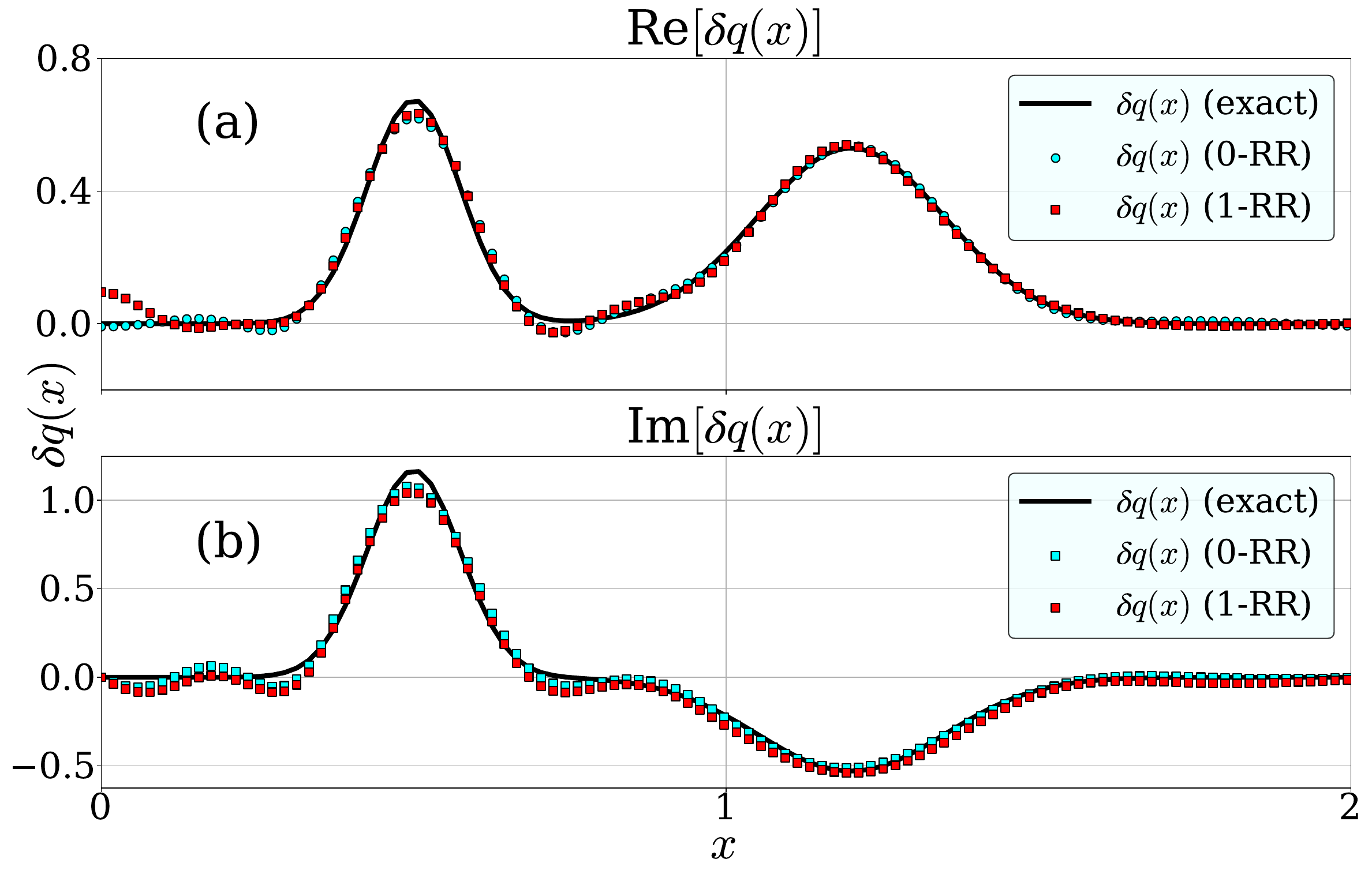}
    \caption{Reconstructing a double Gaussian perturbation $\boldsymbol{\delta q}$ (exact) from theoretical data $\boldsymbol{\delta\lambda}$. Discrete points show inverse problem solution using the ridge regression (RR) based on the first ($n = 0$) and the second ($n = 1$) soliton kernels. Panels (a) and (b) present the real and imaginary parts of the Gaussian perturbation, respectively.}
    \label{Multi_Gaussian_Ex_Rec}
\end{figure}

The LS and RR methods using the fundamental soliton kernel demonstrate high precision reconstruction of both the real and imaginary parts of $\boldsymbol{\delta q}$. The RR method using the second soliton kernel exhibits significant error for $\text{Im}\left[\boldsymbol{\delta q}\right]$, indicating the lack of information obtained from this soliton. In addition we test the inverse reconstruction of more complex perturbations combining two Gaussian perturbations with different parameters. Again the RR method demonstrates high-precision performance, while the LS method is no longer applicable in this case, see Fig.~\ref{Multi_Gaussian_Ex_Rec}.

Now we follow the concept of perturbation reconstruction shown in Fig.~\ref{fig:inv-concept}. We use the same set of $M=16$ perturbed probe pulses $\boldsymbol{q}^{\text{in}}$ and compute their evolution under the NLSE over a time interval of $T = 1.5$ using the standard Split-Step Fourier Method (SSFM). The SSFM evolution is performed in the domain $L = [-30, 30]$, discretized with $N_{\text{SSFM}} = 8192$ points, ensuring that the initial potential decays to machine precision at the domain boundaries. The evolution step is chosen in range $[10^{-6}, 10^{-4}]$ depending on $A$ to maintain the numerical precision.

Once the output potentials $\boldsymbol{q}^{\text{out}}$ have been obtained after the SSFM evolution, we use the Fourier Collocation (FC) method \cite{yang2010nonlinear} to find set of discrete eigenvalues $\boldsymbol{\lambda}_n$ of the first ($n = 0$) and the second ($n=1$) solitons in these potentials. For the FC, the same domain $L = [-30, 30]$ is used, but the wave field is downsampled to $N_{\text{FC}} = 2048$ points; Fourier transforms of $\boldsymbol{q}^{\text{out}}$ are performed using the FFTW library \cite{frigo2005design}. Subtracting the initial unperturbed values $\lambda^{\text{ini}}_n = \iu (A_i - n - 1/2)$, see Eq.~(\ref{Eig_DS}), from the measured ones $\lambda_n$, we finally obtain the desired dataset of eigenvalue shifts $\boldsymbol{\delta \lambda}_n$. Thus, the output of this stage consists of a set of $M = 16$ eigenvalue shifts $\boldsymbol{\delta \lambda}_n$ for the fundamental and second solitons of potentials $\boldsymbol{q}^{\text{out}}$.

\begin{figure}[h!]
    \centering
    \includegraphics[width=1.01\linewidth]{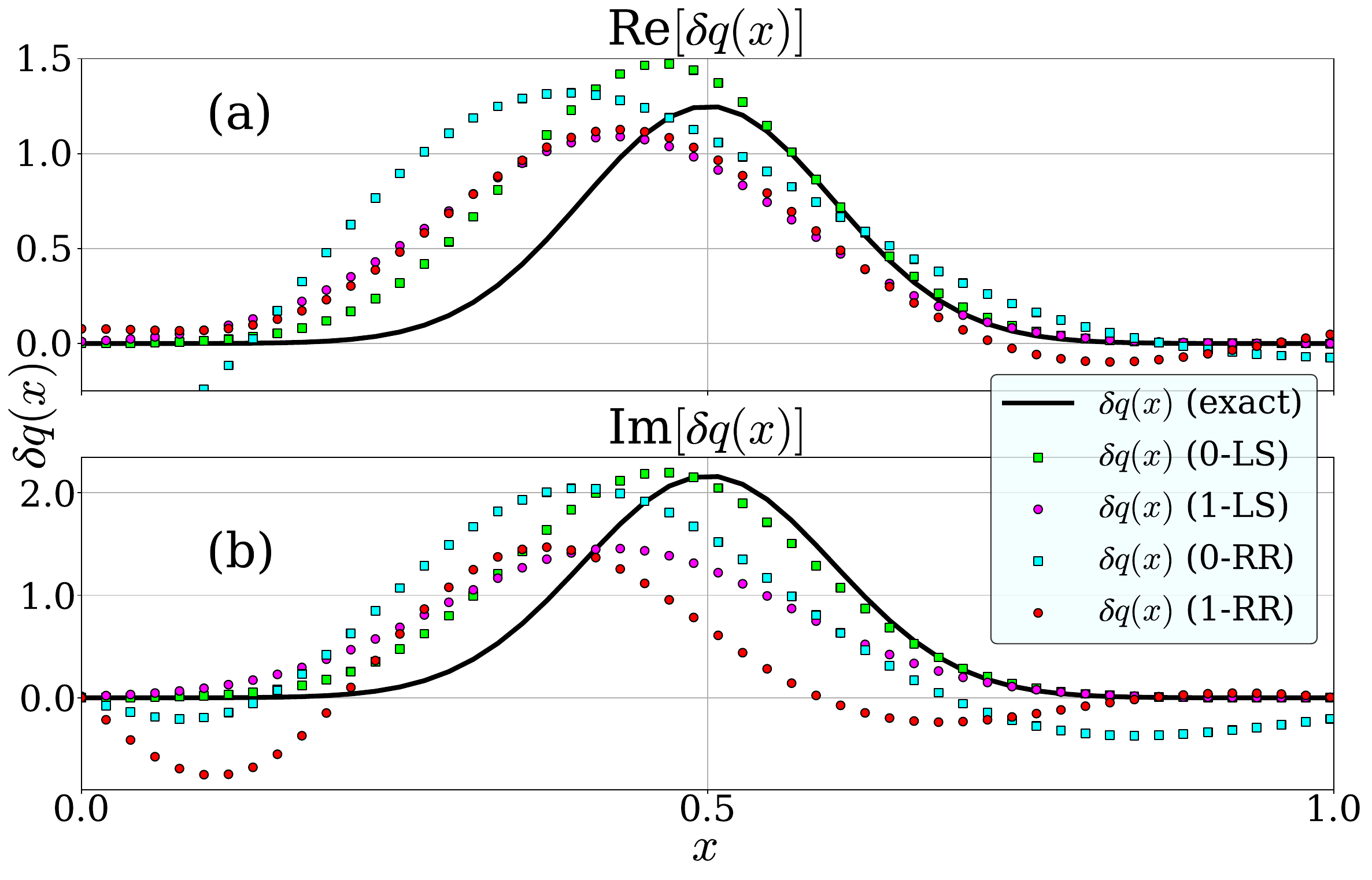}
    \caption{Reconstructing the single Gaussian perturbation (exact) with $a = 2.5, \sigma = 0.1, x_0 = 0.5$, and $\theta = \pi/3$  via the LS and ridge regression using the first ($n = 0$) and the second ($n = 1$) soliton kernels. Left and right panels present the real and imaginary parts of the perturbation, respectively. The relative noise level in the eigenvalues is $\approx 10^{-2}$.}
    \label{Single_Gaussian_Num_Rec}
\end{figure}

\begin{figure}[h!]
    \centering
    \includegraphics[width=1.01\linewidth]{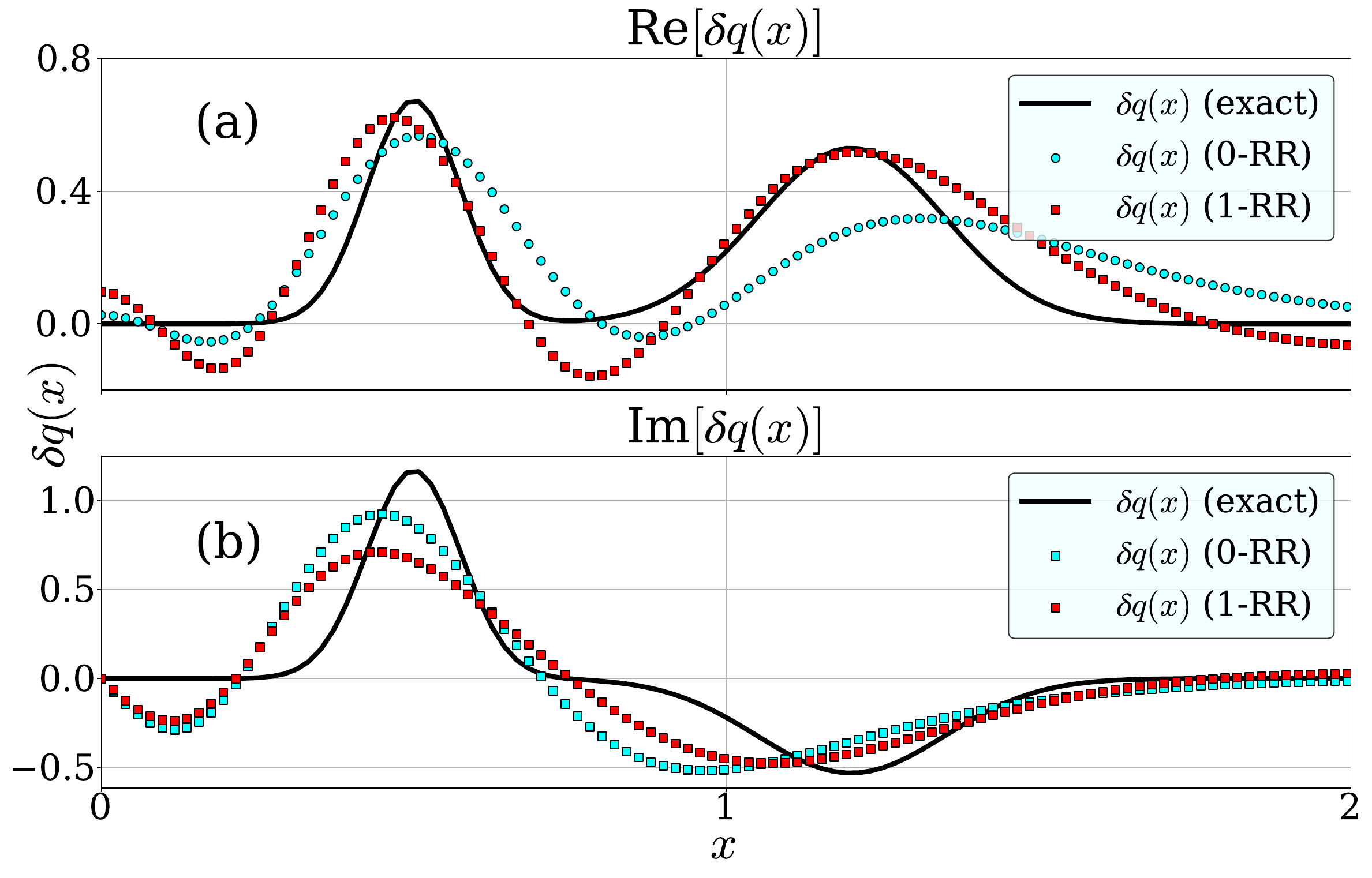}
    \caption{Reconstructing the complex perturbation (exact) using the ridge regression with the first ($n = 0$) and the second ($n = 1$) soliton kernels. Left and right panels present the real and imaginary parts of the Gaussian perturbation, respectively.}
    \label{Multi_Gaussian_Num_Rec}
\end{figure}

Now, with the obtained eigenvalue shifts $\boldsymbol{\delta\lambda}^{\text{num}}_n$, we use again Eqs.~(\ref{dq_reconstruct}) to reconstruct perturbation (\ref{dq_1gauss}) using the LS and RR methods. Regularization parameter $\alpha$ in the RR is now chosen in the range $[10^{-4}, 10^{-3}]$; it should typically be set to the estimated level of deviation between $\boldsymbol{\delta\lambda}^{\text{num}}_n$ and $\boldsymbol{\delta\lambda}^{\text{ex}}_n$. We show the results of single Gaussian perturbation reconstruction in Fig.~\ref{Single_Gaussian_Num_Rec}. The inverse reconstruction of two combined Gaussian perturbations with different parameters is presented in Fig.~\ref{Multi_Gaussian_Num_Rec}. The results remain similar for a wide range of Gaussian parameters $a, \sigma, x_0, \theta$, and even for its different orders (hyper-Gauss), as long as one remains within the applicability of perturbation theory. In Fig. \ref{Single_Gaussian_Num_Rec}, the results obtained using the 0-kernel proved to be better than those obtained using the 1-kernel; however, under different parameters of the perturbation, the situation may reverse. This clearly illustrates the advantage of using two soliton kernels.

Increasing the number of datasets $M$ directly increases the amount of information about the perturbation, enabling more accurate pulse reconstruction; however, the amplitude range of $A_i$ should also be expanded. Increasing $M$ within a fixed amplitude range may, in contrast, lead to worse results, because nearly linearly dependent rows will be continually added to the kernel matrix $\boldsymbol{K}$, thus worsening its condition number. Nevertheless, increasing both the number of datasets $M$ and the amplitude range $A_i$ simultaneously can enhance the resolution of the pulse reconstruction. This is particularly relevant when considering any rather narrow high-frequency pulses with a broad Fourier spectrum. Higher amplitudes make the RFs narrower, enabling them to sense higher-frequency components more strongly. Thus, selecting the key parameters involves trade-offs and depends on the specifics of the particular problem.

Additionally, we test our approach by incorporating small random fluctuations $\delta$ into the eigenvalue shifts $\delta \lambda_n$, thereby simulating noise and errors of various sources (such as uncertainties in initial signal amplitudes, non-sech pulse shapes, etc.). Note that the maximum noise-induced effects must not exceed those of the perturbation itself. In our simulations, the proposed method is shown to be effective under random fluctuations of the order of $\delta \approx 10^{-3}$.

\section{Conclusions}
Our work advances the recently proposed approach of soliton eigenvalue response functions \cite{mucci2025manipulation} and suggests using the fundamental integrability property of the NLSE for inverse sensing problems in nonlinear media and materials. Taking advantage of the exact solvability of the ZS system for the sech-potential, we have derived analytical expressions for the RFs and demonstrated efficient soliton manipulation, which can be experimentally verified in different physical systems. For example, as done for box-shaped pulses in a recirculating optical fiber loop setup with a CW drive and optical modulators \cite{mucci2025manipulation}. Alternatively, manipulation of strongly interacting solitons using the eigenvalue RFs can be achieved in a long-distance water-wave tank with wave makers to control the shape of the fluid surface \cite{chabchoub2012super,chabchoub2012observation}, or in other nonlinear systems that enable control of the injected pulse shapes. Being close to common experimental pulse profiles, the sech-shaped waveforms considered analytically here provide deep insight into the general behavior of the RFs, which, when necessary, can be computed numerically for specific pulse shapes; see Fig. 1, where we demonstrated numerical reproduction of the analytically obtained results. Even computed numerically, the RFs allow semi-analytical analysis of the eigenvalue manipulation task thanks to the overlap integral interpretation of the perturbation action provided by Eq.~(\ref{Eigs_RF}), which is the key feature compared to other approaches to splitting multi-soliton pulses into individual solitons and soliton eigenvalue manipulations \cite{satsuma1974b,kaup1994effect,prilepsky2007breakup,gao2026soliton}.

In the second part of the work we propose an approach to sensing instant pulse perturbations, that leverages the nonlinear propagation of the probe signal over an unknown distance, thereby enabling the extraction of information about the perturbation source hidden within nonlinear media or materials. We assume that the amplitude is a known varying parameter of the probe pulses, yielding a data set of eigenvalue deviations as a function of $A$ expressed via the overlap integrals (\ref{Eigs_RF_A}), which allows us to formulate the inverse problem for perturbation reconstruction. Considering a general situation, when the number of data sets $M$ obtained at different probe signal amplitudes is different from the number of signal discretization points $N$, we demonstrate robust solving of the inverse problem using the pseudo-inverse matrix of $M\times N $ size regularized using the ridge regression. Remarkably, a small number of measured transmitted pulses $M \ll N$ is sufficient to reconstruct a Gaussian and more complex-shaped perturbations.

We tested the noise stability of the inverse problem solution and demonstrated that robust reconstruction of probe signal perturbations is possible for soliton eigenvalue distortions at the level up to $10^{-3}$. Verifying a realistic scenario for the propagation of probe signals in nonlinear media is the subject of a separate study and requires estimates of typical noise levels for the chosen optical, hydrodynamic, or other physical testbeds. Here are two advantages of the eigenvalue RFs approach that might help improve the robustness of solutions to the inverse problem. First, using discrete soliton eigenvalues as a source of information filters the probe signal noise, leaving only specific harmonics which are able to translate into the chosen soliton, see e.g. \cite{mullyadzhanov2021solitons}. Second, repeated propagation of the same probe pulses can yield averaged eigenvalue changes $\left<\delta\lambda_n(A_i)\right>$ with significantly reduced noise-induced corrections, enabling a more robust pulse perturbations reconstruction.

The described inverse problem for soliton eigenvalue RFs can be used to reconstruct high-intensity perturbations that can alter high-amplitude soliton signals. In contrast to conventional linear remote sensing approaches \cite{elachi2021introduction}, using soliton signals is justified for highly nonlinear media or long-distance nonlinear systems. Previously, nonlinear second-harmonic generation and solitons have been used for imaging microscopy, optical tomography, and acoustic detection of defects and strains in various materials \cite{campagnola2003second,hu2020harmonic,hong2002nondestructive,khatri2008highly,li2013solitary,jalali2022detection,yoon2023defect}. The major advantage of our approach is that it provides sufficient information to fully solve the inverse problem and reconstruct the shape of the perturbation when direct measurements are unavailable.

Our work is the first step towards soliton tomography of perturbations or media distortions distributed along a nonlinear propagation channel. The evolution of multi-soliton probe signals can fingerprint the location of the distortion in the eigenvalue portrait, as it does for the shape of instant perturbations. The development of a complete soliton tomography approach requires the numerical computation of spatio-temporal soliton-eigenvalue response functions $\mathrm{RF}^{\mathrm{Re/Im}}(x,t)$. Additional information to strengthen the robustness of the inverse problem can be obtained from changes in soliton phases, modifications of the reflection coefficient, and asymmetric probe pulses, which offer many avenues for bringing the presented soliton eigenvalue sensing concept to real-world physical applications.

\begin{acknowledgments}
This work was supported by the Russian Science Foundation project N\textsuperscript{\underline{\scriptsize o}} 25-72-31023.
\end{acknowledgments}

%%%%%%%%%%%%%%%%%
%%%%% BIBLIOGRAPHY
%%%%%%%%%%%%%%%%%
%\bibliographystyle{apsrev4-1}
%\bibliography{RFs-approach}

\begin{thebibliography}{47}%
\makeatletter
\providecommand \@ifxundefined [1]{%
 \@ifx{#1\undefined}
}%
\providecommand \@ifnum [1]{%
 \ifnum #1\expandafter \@firstoftwo
 \else \expandafter \@secondoftwo
 \fi
}%
\providecommand \@ifx [1]{%
 \ifx #1\expandafter \@firstoftwo
 \else \expandafter \@secondoftwo
 \fi
}%
\providecommand \natexlab [1]{#1}%
\providecommand \enquote  [1]{``#1''}%
\providecommand \bibnamefont  [1]{#1}%
\providecommand \bibfnamefont [1]{#1}%
\providecommand \citenamefont [1]{#1}%
\providecommand \href@noop [0]{\@secondoftwo}%
\providecommand \href [0]{\begingroup \@sanitize@url \@href}%
\providecommand \@href[1]{\@@startlink{#1}\@@href}%
\providecommand \@@href[1]{\endgroup#1\@@endlink}%
\providecommand \@sanitize@url [0]{\catcode `\\12\catcode `\$12\catcode
  `\&12\catcode `\#12\catcode `\^12\catcode `\_12\catcode `\%12\relax}%
\providecommand \@@startlink[1]{}%
\providecommand \@@endlink[0]{}%
\providecommand \url  [0]{\begingroup\@sanitize@url \@url }%
\providecommand \@url [1]{\endgroup\@href {#1}{\urlprefix }}%
\providecommand \urlprefix  [0]{URL }%
\providecommand \Eprint [0]{\href }%
\providecommand \doibase [0]{http://dx.doi.org/}%
\providecommand \selectlanguage [0]{\@gobble}%
\providecommand \bibinfo  [0]{\@secondoftwo}%
\providecommand \bibfield  [0]{\@secondoftwo}%
\providecommand \translation [1]{[#1]}%
\providecommand \BibitemOpen [0]{}%
\providecommand \bibitemStop [0]{}%
\providecommand \bibitemNoStop [0]{.\EOS\space}%
\providecommand \EOS [0]{\spacefactor3000\relax}%
\providecommand \BibitemShut  [1]{\csname bibitem#1\endcsname}%
\let\auto@bib@innerbib\@empty
%</preamble>
\bibitem [{\citenamefont {Zakharov}\ and\ \citenamefont
  {Shabat}(1972)}]{zakharov1972exact}%
  \BibitemOpen
  \bibfield  {author} {\bibinfo {author} {\bibfnamefont {V.~E.}\ \bibnamefont
  {Zakharov}}\ and\ \bibinfo {author} {\bibfnamefont {A.~B.}\ \bibnamefont
  {Shabat}},\ }\href@noop {} {\bibfield  {journal} {\bibinfo  {journal} {Sov.
  Phys. JETP}\ }\textbf {\bibinfo {volume} {34}},\ \bibinfo {pages} {62}
  (\bibinfo {year} {1972})}\BibitemShut {NoStop}%
\bibitem [{\citenamefont {Novikov}\ \emph {et~al.}(1984)\citenamefont
  {Novikov}, \citenamefont {Manakov}, \citenamefont {Pitaevskii},\ and\
  \citenamefont {Zakharov}}]{NovikovBook1984}%
  \BibitemOpen
  \bibfield  {author} {\bibinfo {author} {\bibfnamefont {S.}~\bibnamefont
  {Novikov}}, \bibinfo {author} {\bibfnamefont {S.}~\bibnamefont {Manakov}},
  \bibinfo {author} {\bibfnamefont {L.}~\bibnamefont {Pitaevskii}}, \ and\
  \bibinfo {author} {\bibfnamefont {V.}~\bibnamefont {Zakharov}},\ }\href@noop
  {} {\emph {\bibinfo {title} {Theory of solitons: the inverse scattering
  method}}}\ (\bibinfo  {publisher} {Springer Science \& Business Media},\
  \bibinfo {year} {1984})\BibitemShut {NoStop}%
\bibitem [{\citenamefont {Ablowitz}\ and\ \citenamefont
  {Segur}(1981)}]{AblowitzBook1981}%
  \BibitemOpen
  \bibfield  {author} {\bibinfo {author} {\bibfnamefont {M.~J.}\ \bibnamefont
  {Ablowitz}}\ and\ \bibinfo {author} {\bibfnamefont {H.}~\bibnamefont
  {Segur}},\ }\href@noop {} {\emph {\bibinfo {title} {Solitons and the inverse
  scattering transform}}},\ Vol.~\bibinfo {volume} {4}\ (\bibinfo  {publisher}
  {Siam},\ \bibinfo {year} {1981})\BibitemShut {NoStop}%
\bibitem [{\citenamefont {Copie}\ \emph {et~al.}(2025)\citenamefont {Copie},
  \citenamefont {Suret},\ and\ \citenamefont {Randoux}}]{copie2025controlled}%
  \BibitemOpen
  \bibfield  {author} {\bibinfo {author} {\bibfnamefont {F.}~\bibnamefont
  {Copie}}, \bibinfo {author} {\bibfnamefont {P.}~\bibnamefont {Suret}}, \ and\
  \bibinfo {author} {\bibfnamefont {S.}~\bibnamefont {Randoux}},\ }\href@noop
  {} {\bibfield  {journal} {\bibinfo  {journal} {Physical Review Research}\
  }\textbf {\bibinfo {volume} {7}},\ \bibinfo {pages} {043232} (\bibinfo {year}
  {2025})}\BibitemShut {NoStop}%
\bibitem [{\citenamefont {Xu}\ \emph {et~al.}(2019)\citenamefont {Xu},
  \citenamefont {Gelash}, \citenamefont {Chabchoub}, \citenamefont {Zakharov},\
  and\ \citenamefont {Kibler}}]{xu2019breather}%
  \BibitemOpen
  \bibfield  {author} {\bibinfo {author} {\bibfnamefont {G.}~\bibnamefont
  {Xu}}, \bibinfo {author} {\bibfnamefont {A.}~\bibnamefont {Gelash}}, \bibinfo
  {author} {\bibfnamefont {A.}~\bibnamefont {Chabchoub}}, \bibinfo {author}
  {\bibfnamefont {V.}~\bibnamefont {Zakharov}}, \ and\ \bibinfo {author}
  {\bibfnamefont {B.}~\bibnamefont {Kibler}},\ }\href@noop {} {\bibfield
  {journal} {\bibinfo  {journal} {Physical review letters}\ }\textbf {\bibinfo
  {volume} {122}},\ \bibinfo {pages} {084101} (\bibinfo {year}
  {2019})}\BibitemShut {NoStop}%
\bibitem [{\citenamefont {Kharif}\ \emph {et~al.}(2008)\citenamefont {Kharif},
  \citenamefont {Pelinovsky},\ and\ \citenamefont
  {Slunyaev}}]{kharif2008rogue}%
  \BibitemOpen
  \bibfield  {author} {\bibinfo {author} {\bibfnamefont {C.}~\bibnamefont
  {Kharif}}, \bibinfo {author} {\bibfnamefont {E.}~\bibnamefont {Pelinovsky}},
  \ and\ \bibinfo {author} {\bibfnamefont {A.}~\bibnamefont {Slunyaev}},\
  }\href@noop {} {\emph {\bibinfo {title} {Rogue waves in the ocean}}}\
  (\bibinfo  {publisher} {Springer Science \& Business Media},\ \bibinfo {year}
  {2008})\BibitemShut {NoStop}%
\bibitem [{\citenamefont {Akhmediev}\ \emph {et~al.}(2009)\citenamefont
  {Akhmediev}, \citenamefont {Ankiewicz},\ and\ \citenamefont
  {Soto-Crespo}}]{akhmediev2009rogue}%
  \BibitemOpen
  \bibfield  {author} {\bibinfo {author} {\bibfnamefont {N.}~\bibnamefont
  {Akhmediev}}, \bibinfo {author} {\bibfnamefont {A.}~\bibnamefont
  {Ankiewicz}}, \ and\ \bibinfo {author} {\bibfnamefont {J.~M.}\ \bibnamefont
  {Soto-Crespo}},\ }\href@noop {} {\bibfield  {journal} {\bibinfo  {journal}
  {Physical Review E—Statistical, Nonlinear, and Soft Matter Physics}\
  }\textbf {\bibinfo {volume} {80}},\ \bibinfo {pages} {026601} (\bibinfo
  {year} {2009})}\BibitemShut {NoStop}%
\bibitem [{\citenamefont {Yousefi}\ and\ \citenamefont
  {Kschischang}(2014)}]{yousefi2014information}%
  \BibitemOpen
  \bibfield  {author} {\bibinfo {author} {\bibfnamefont {M.~I.}\ \bibnamefont
  {Yousefi}}\ and\ \bibinfo {author} {\bibfnamefont {F.~R.}\ \bibnamefont
  {Kschischang}},\ }\href@noop {} {\bibfield  {journal} {\bibinfo  {journal}
  {IEEE Transactions on Information Theory}\ }\textbf {\bibinfo {volume}
  {60}},\ \bibinfo {pages} {4312} (\bibinfo {year} {2014})}\BibitemShut
  {NoStop}%
\bibitem [{\citenamefont {Turitsyn}\ \emph {et~al.}(2017)\citenamefont
  {Turitsyn}, \citenamefont {Prilepsky}, \citenamefont {Le}, \citenamefont
  {Wahls}, \citenamefont {Frumin}, \citenamefont {Kamalian},\ and\
  \citenamefont {Derevyanko}}]{turitsyn2017nonlinear}%
  \BibitemOpen
  \bibfield  {author} {\bibinfo {author} {\bibfnamefont {S.~K.}\ \bibnamefont
  {Turitsyn}}, \bibinfo {author} {\bibfnamefont {J.~E.}\ \bibnamefont
  {Prilepsky}}, \bibinfo {author} {\bibfnamefont {S.~T.}\ \bibnamefont {Le}},
  \bibinfo {author} {\bibfnamefont {S.}~\bibnamefont {Wahls}}, \bibinfo
  {author} {\bibfnamefont {L.~L.}\ \bibnamefont {Frumin}}, \bibinfo {author}
  {\bibfnamefont {M.}~\bibnamefont {Kamalian}}, \ and\ \bibinfo {author}
  {\bibfnamefont {S.~A.}\ \bibnamefont {Derevyanko}},\ }\href@noop {}
  {\bibfield  {journal} {\bibinfo  {journal} {Optica}\ }\textbf {\bibinfo
  {volume} {4}},\ \bibinfo {pages} {307} (\bibinfo {year} {2017})}\BibitemShut
  {NoStop}%
\bibitem [{\citenamefont {Le}\ \emph {et~al.}(2017)\citenamefont {Le},
  \citenamefont {Aref},\ and\ \citenamefont {Buelow}}]{le2017nonlinear}%
  \BibitemOpen
  \bibfield  {author} {\bibinfo {author} {\bibfnamefont {S.~T.}\ \bibnamefont
  {Le}}, \bibinfo {author} {\bibfnamefont {V.}~\bibnamefont {Aref}}, \ and\
  \bibinfo {author} {\bibfnamefont {H.}~\bibnamefont {Buelow}},\ }\href@noop {}
  {\bibfield  {journal} {\bibinfo  {journal} {Nature Photonics}\ }\textbf
  {\bibinfo {volume} {11}},\ \bibinfo {pages} {570} (\bibinfo {year}
  {2017})}\BibitemShut {NoStop}%
\bibitem [{\citenamefont {Karpman}\ and\ \citenamefont
  {Maslov}(1977)}]{KarpmanJETP1977}%
  \BibitemOpen
  \bibfield  {author} {\bibinfo {author} {\bibfnamefont {V.~I.}\ \bibnamefont
  {Karpman}}\ and\ \bibinfo {author} {\bibfnamefont {E.~M.}\ \bibnamefont
  {Maslov}},\ }\href@noop {} {\bibfield  {journal} {\bibinfo  {journal} {Soviet
  Physics JETP}\ }\textbf {\bibinfo {volume} {46}},\ \bibinfo {pages} {281}
  (\bibinfo {year} {1977})}\BibitemShut {NoStop}%
\bibitem [{\citenamefont {Kaup}(1976)}]{KaupSIAM1976}%
  \BibitemOpen
  \bibfield  {author} {\bibinfo {author} {\bibfnamefont {D.~J.}\ \bibnamefont
  {Kaup}},\ }\href@noop {} {\bibfield  {journal} {\bibinfo  {journal} {SIAM
  Journal on Applied Mathematics}\ }\textbf {\bibinfo {volume} {31}},\ \bibinfo
  {pages} {121} (\bibinfo {year} {1976})}\BibitemShut {NoStop}%
\bibitem [{\citenamefont {Kivshar}\ and\ \citenamefont
  {Malomed}(1989)}]{KivsharRMP1989}%
  \BibitemOpen
  \bibfield  {author} {\bibinfo {author} {\bibfnamefont {Y.~S.}\ \bibnamefont
  {Kivshar}}\ and\ \bibinfo {author} {\bibfnamefont {B.~A.}\ \bibnamefont
  {Malomed}},\ }\href@noop {} {\bibfield  {journal} {\bibinfo  {journal}
  {Reviews of Modern Physics}\ }\textbf {\bibinfo {volume} {61}},\ \bibinfo
  {pages} {763} (\bibinfo {year} {1989})}\BibitemShut {NoStop}%
\bibitem [{\citenamefont {Osborne}(2010)}]{OsborneBook2010}%
  \BibitemOpen
  \bibfield  {author} {\bibinfo {author} {\bibfnamefont {A.}~\bibnamefont
  {Osborne}},\ }\href@noop {} {\emph {\bibinfo {title} {{Nonlinear ocean
  waves}}}}\ (\bibinfo  {publisher} {Academic Press},\ \bibinfo {year}
  {2010})\BibitemShut {NoStop}%
\bibitem [{\citenamefont {Wahls}\ and\ \citenamefont
  {Poor}(2015)}]{wahls2015fast}%
  \BibitemOpen
  \bibfield  {author} {\bibinfo {author} {\bibfnamefont {S.}~\bibnamefont
  {Wahls}}\ and\ \bibinfo {author} {\bibfnamefont {H.~V.}\ \bibnamefont
  {Poor}},\ }\href@noop {} {\bibfield  {journal} {\bibinfo  {journal} {IEEE
  Transactions on Information Theory}\ }\textbf {\bibinfo {volume} {61}},\
  \bibinfo {pages} {6957} (\bibinfo {year} {2015})}\BibitemShut {NoStop}%
\bibitem [{\citenamefont {Gelash}\ and\ \citenamefont
  {Agafontsev}(2018)}]{gelash2018strongly}%
  \BibitemOpen
  \bibfield  {author} {\bibinfo {author} {\bibfnamefont {A.}~\bibnamefont
  {Gelash}}\ and\ \bibinfo {author} {\bibfnamefont {D.}~\bibnamefont
  {Agafontsev}},\ }\href@noop {} {\bibfield  {journal} {\bibinfo  {journal}
  {Physical Review E}\ }\textbf {\bibinfo {volume} {98}},\ \bibinfo {pages}
  {042210} (\bibinfo {year} {2018})}\BibitemShut {NoStop}%
\bibitem [{\citenamefont {Suret}\ \emph {et~al.}(2020)\citenamefont {Suret},
  \citenamefont {Tikan}, \citenamefont {Bonnefoy}, \citenamefont {Copie},
  \citenamefont {Ducrozet}, \citenamefont {Gelash}, \citenamefont
  {Prabhudesai}, \citenamefont {Michel}, \citenamefont {Cazaubiel},
  \citenamefont {Falcon} \emph {et~al.}}]{suret2020nonlinear}%
  \BibitemOpen
  \bibfield  {author} {\bibinfo {author} {\bibfnamefont {P.}~\bibnamefont
  {Suret}}, \bibinfo {author} {\bibfnamefont {A.}~\bibnamefont {Tikan}},
  \bibinfo {author} {\bibfnamefont {F.}~\bibnamefont {Bonnefoy}}, \bibinfo
  {author} {\bibfnamefont {F.}~\bibnamefont {Copie}}, \bibinfo {author}
  {\bibfnamefont {G.}~\bibnamefont {Ducrozet}}, \bibinfo {author}
  {\bibfnamefont {A.}~\bibnamefont {Gelash}}, \bibinfo {author} {\bibfnamefont
  {G.}~\bibnamefont {Prabhudesai}}, \bibinfo {author} {\bibfnamefont
  {G.}~\bibnamefont {Michel}}, \bibinfo {author} {\bibfnamefont
  {A.}~\bibnamefont {Cazaubiel}}, \bibinfo {author} {\bibfnamefont
  {E.}~\bibnamefont {Falcon}},  \emph {et~al.},\ }\href@noop {} {\bibfield
  {journal} {\bibinfo  {journal} {Physical Review Letters}\ }\textbf {\bibinfo
  {volume} {125}},\ \bibinfo {pages} {264101} (\bibinfo {year}
  {2020})}\BibitemShut {NoStop}%
\bibitem [{\citenamefont {Fache}\ \emph {et~al.}(2025)\citenamefont {Fache},
  \citenamefont {Damart}, \citenamefont {Copie}, \citenamefont {Bonnemain},
  \citenamefont {Congy}, \citenamefont {Roberti}, \citenamefont {Suret},
  \citenamefont {El},\ and\ \citenamefont {Randoux}}]{fache2025dissipation}%
  \BibitemOpen
  \bibfield  {author} {\bibinfo {author} {\bibfnamefont {L.}~\bibnamefont
  {Fache}}, \bibinfo {author} {\bibfnamefont {H.}~\bibnamefont {Damart}},
  \bibinfo {author} {\bibfnamefont {F.}~\bibnamefont {Copie}}, \bibinfo
  {author} {\bibfnamefont {T.}~\bibnamefont {Bonnemain}}, \bibinfo {author}
  {\bibfnamefont {T.}~\bibnamefont {Congy}}, \bibinfo {author} {\bibfnamefont
  {G.}~\bibnamefont {Roberti}}, \bibinfo {author} {\bibfnamefont
  {P.}~\bibnamefont {Suret}}, \bibinfo {author} {\bibfnamefont
  {G.}~\bibnamefont {El}}, \ and\ \bibinfo {author} {\bibfnamefont
  {S.}~\bibnamefont {Randoux}},\ }\href@noop {} {\bibfield  {journal} {\bibinfo
   {journal} {Physical Review Letters}\ }\textbf {\bibinfo {volume} {134}},\
  \bibinfo {pages} {147201} (\bibinfo {year} {2025})}\BibitemShut {NoStop}%
\bibitem [{\citenamefont {Mucci}\ \emph {et~al.}(2025)\citenamefont {Mucci},
  \citenamefont {Suret}, \citenamefont {Copie}, \citenamefont {Randoux},
  \citenamefont {Mullyadzhanov},\ and\ \citenamefont
  {Gelash}}]{mucci2025manipulation}%
  \BibitemOpen
  \bibfield  {author} {\bibinfo {author} {\bibfnamefont {A.}~\bibnamefont
  {Mucci}}, \bibinfo {author} {\bibfnamefont {P.}~\bibnamefont {Suret}},
  \bibinfo {author} {\bibfnamefont {F.}~\bibnamefont {Copie}}, \bibinfo
  {author} {\bibfnamefont {S.}~\bibnamefont {Randoux}}, \bibinfo {author}
  {\bibfnamefont {R.}~\bibnamefont {Mullyadzhanov}}, \ and\ \bibinfo {author}
  {\bibfnamefont {A.}~\bibnamefont {Gelash}},\ }\href@noop {} {\bibfield
  {journal} {\bibinfo  {journal} {Physical Review Letters}\ }\textbf {\bibinfo
  {volume} {134}},\ \bibinfo {pages} {193804} (\bibinfo {year}
  {2025})}\BibitemShut {NoStop}%
\bibitem [{\citenamefont {Landau}\ and\ \citenamefont
  {Lifshitz}(1958)}]{landau1958quantum}%
  \BibitemOpen
  \bibfield  {author} {\bibinfo {author} {\bibfnamefont {L.~D.}\ \bibnamefont
  {Landau}}\ and\ \bibinfo {author} {\bibfnamefont {E.~M.}\ \bibnamefont
  {Lifshitz}},\ }\href@noop {} {\emph {\bibinfo {title} {{Quantum Mechanics:
  Non-relativistic Theory. V. 3 of Course of Theoretical Physics}}}}\ (\bibinfo
   {publisher} {Pergamon Press},\ \bibinfo {year} {1958})\BibitemShut {NoStop}%
\bibitem [{\citenamefont {Satsuma}\ and\ \citenamefont
  {Yajima}(1974)}]{satsuma1974b}%
  \BibitemOpen
  \bibfield  {author} {\bibinfo {author} {\bibfnamefont {J.}~\bibnamefont
  {Satsuma}}\ and\ \bibinfo {author} {\bibfnamefont {N.}~\bibnamefont
  {Yajima}},\ }\href@noop {} {\bibfield  {journal} {\bibinfo  {journal}
  {Progress of Theoretical Physics Supplement}\ }\textbf {\bibinfo {volume}
  {55}},\ \bibinfo {pages} {284} (\bibinfo {year} {1974})}\BibitemShut
  {NoStop}%
\bibitem [{\citenamefont {Boffetta}\ and\ \citenamefont
  {Osborne}(1992)}]{boffetta1992computation}%
  \BibitemOpen
  \bibfield  {author} {\bibinfo {author} {\bibfnamefont {G.}~\bibnamefont
  {Boffetta}}\ and\ \bibinfo {author} {\bibfnamefont {A.~R.}\ \bibnamefont
  {Osborne}},\ }\href@noop {} {\bibfield  {journal} {\bibinfo  {journal}
  {Journal of Computational Physics}\ }\textbf {\bibinfo {volume} {102}},\
  \bibinfo {pages} {252} (\bibinfo {year} {1992})}\BibitemShut {NoStop}%
\bibitem [{\citenamefont {Mullyadzhanov}\ and\ \citenamefont
  {Gelash}(2019)}]{mullyadzhanov2019direct}%
  \BibitemOpen
  \bibfield  {author} {\bibinfo {author} {\bibfnamefont {R.}~\bibnamefont
  {Mullyadzhanov}}\ and\ \bibinfo {author} {\bibfnamefont {A.}~\bibnamefont
  {Gelash}},\ }\href@noop {} {\bibfield  {journal} {\bibinfo  {journal} {Optics
  letters}\ }\textbf {\bibinfo {volume} {44}},\ \bibinfo {pages} {5298}
  (\bibinfo {year} {2019})}\BibitemShut {NoStop}%
\bibitem [{\citenamefont {Medvedev}\ \emph {et~al.}(2024)\citenamefont
  {Medvedev}, \citenamefont {Vaseva}, \citenamefont {Kachulin}, \citenamefont
  {Chekhovskoy},\ and\ \citenamefont {Fedoruk}}]{medvedev2024fast}%
  \BibitemOpen
  \bibfield  {author} {\bibinfo {author} {\bibfnamefont {S.}~\bibnamefont
  {Medvedev}}, \bibinfo {author} {\bibfnamefont {I.}~\bibnamefont {Vaseva}},
  \bibinfo {author} {\bibfnamefont {D.}~\bibnamefont {Kachulin}}, \bibinfo
  {author} {\bibfnamefont {I.}~\bibnamefont {Chekhovskoy}}, \ and\ \bibinfo
  {author} {\bibfnamefont {M.}~\bibnamefont {Fedoruk}},\ }\href@noop {}
  {\bibfield  {journal} {\bibinfo  {journal} {Optics Letters}\ }\textbf
  {\bibinfo {volume} {49}},\ \bibinfo {pages} {1884} (\bibinfo {year}
  {2024})}\BibitemShut {NoStop}%
\bibitem [{\citenamefont {Gelash}\ and\ \citenamefont
  {Mullyadzhanov}(2020)}]{Gelash2020}%
  \BibitemOpen
  \bibfield  {author} {\bibinfo {author} {\bibfnamefont {A.}~\bibnamefont
  {Gelash}}\ and\ \bibinfo {author} {\bibfnamefont {R.}~\bibnamefont
  {Mullyadzhanov}},\ }\href@noop {} {\bibfield  {journal} {\bibinfo  {journal}
  {Physical Review E}\ }\textbf {\bibinfo {volume} {101}},\ \bibinfo {pages}
  {052206} (\bibinfo {year} {2020})}\BibitemShut {NoStop}%
\bibitem [{\citenamefont {Mullyadzhanov}\ and\ \citenamefont
  {Gelash}(2021)}]{mullyadzhanov2021solitons}%
  \BibitemOpen
  \bibfield  {author} {\bibinfo {author} {\bibfnamefont {R.}~\bibnamefont
  {Mullyadzhanov}}\ and\ \bibinfo {author} {\bibfnamefont {A.}~\bibnamefont
  {Gelash}},\ }\href@noop {} {\bibfield  {journal} {\bibinfo  {journal}
  {Physical Review Letters}\ }\textbf {\bibinfo {volume} {126}},\ \bibinfo
  {pages} {234101} (\bibinfo {year} {2021})}\BibitemShut {NoStop}%
\bibitem [{\citenamefont {Tai}\ \emph {et~al.}(1988)\citenamefont {Tai},
  \citenamefont {Hasegawa},\ and\ \citenamefont {Bekki}}]{tai1988fission}%
  \BibitemOpen
  \bibfield  {author} {\bibinfo {author} {\bibfnamefont {K.}~\bibnamefont
  {Tai}}, \bibinfo {author} {\bibfnamefont {A.}~\bibnamefont {Hasegawa}}, \
  and\ \bibinfo {author} {\bibfnamefont {N.}~\bibnamefont {Bekki}},\
  }\href@noop {} {\bibfield  {journal} {\bibinfo  {journal} {Optics letters}\
  }\textbf {\bibinfo {volume} {13}},\ \bibinfo {pages} {392} (\bibinfo {year}
  {1988})}\BibitemShut {NoStop}%
\bibitem [{\citenamefont {Kaup}\ \emph {et~al.}(1994)\citenamefont {Kaup},
  \citenamefont {El-Reedy},\ and\ \citenamefont {Malomed}}]{kaup1994effect}%
  \BibitemOpen
  \bibfield  {author} {\bibinfo {author} {\bibfnamefont {D.}~\bibnamefont
  {Kaup}}, \bibinfo {author} {\bibfnamefont {J.}~\bibnamefont {El-Reedy}}, \
  and\ \bibinfo {author} {\bibfnamefont {B.~A.}\ \bibnamefont {Malomed}},\
  }\href@noop {} {\bibfield  {journal} {\bibinfo  {journal} {Physical Review
  E}\ }\textbf {\bibinfo {volume} {50}},\ \bibinfo {pages} {1635} (\bibinfo
  {year} {1994})}\BibitemShut {NoStop}%
\bibitem [{\citenamefont {Prilepsky}\ and\ \citenamefont
  {Derevyanko}(2007)}]{prilepsky2007breakup}%
  \BibitemOpen
  \bibfield  {author} {\bibinfo {author} {\bibfnamefont {J.~E.}\ \bibnamefont
  {Prilepsky}}\ and\ \bibinfo {author} {\bibfnamefont {S.~A.}\ \bibnamefont
  {Derevyanko}},\ }\href@noop {} {\bibfield  {journal} {\bibinfo  {journal}
  {Physical Review E—Statistical, Nonlinear, and Soft Matter Physics}\
  }\textbf {\bibinfo {volume} {75}},\ \bibinfo {pages} {036616} (\bibinfo
  {year} {2007})}\BibitemShut {NoStop}%
\bibitem [{\citenamefont {Gao}\ \emph {et~al.}(2026)\citenamefont {Gao},
  \citenamefont {Wang}, \citenamefont {An}, \citenamefont {Wen}, \citenamefont
  {Zheng}, \citenamefont {Li},\ and\ \citenamefont {Gao}}]{gao2026soliton}%
  \BibitemOpen
  \bibfield  {author} {\bibinfo {author} {\bibfnamefont {P.}~\bibnamefont
  {Gao}}, \bibinfo {author} {\bibfnamefont {X.}~\bibnamefont {Wang}}, \bibinfo
  {author} {\bibfnamefont {S.}~\bibnamefont {An}}, \bibinfo {author}
  {\bibfnamefont {K.}~\bibnamefont {Wen}}, \bibinfo {author} {\bibfnamefont
  {J.}~\bibnamefont {Zheng}}, \bibinfo {author} {\bibfnamefont
  {T.}~\bibnamefont {Li}}, \ and\ \bibinfo {author} {\bibfnamefont
  {P.}~\bibnamefont {Gao}},\ }\href@noop {} {\bibfield  {journal} {\bibinfo
  {journal} {Optics Express}\ }\textbf {\bibinfo {volume} {34}},\ \bibinfo
  {pages} {9925} (\bibinfo {year} {2026})}\BibitemShut {NoStop}%
\bibitem [{\citenamefont {Doicu}\ \emph {et~al.}(2010)\citenamefont {Doicu},
  \citenamefont {Trautmann},\ and\ \citenamefont
  {Schreier}}]{doicu2010numerical}%
  \BibitemOpen
  \bibfield  {author} {\bibinfo {author} {\bibfnamefont {A.}~\bibnamefont
  {Doicu}}, \bibinfo {author} {\bibfnamefont {T.}~\bibnamefont {Trautmann}}, \
  and\ \bibinfo {author} {\bibfnamefont {F.}~\bibnamefont {Schreier}},\
  }\href@noop {} {\emph {\bibinfo {title} {Numerical regularization for
  atmospheric inverse problems}}}\ (\bibinfo  {publisher} {Springer Science \&
  Business Media},\ \bibinfo {year} {2010})\BibitemShut {NoStop}%
\bibitem [{\citenamefont {Axler}(2024)}]{axler2024linear}%
  \BibitemOpen
  \bibfield  {author} {\bibinfo {author} {\bibfnamefont {S.}~\bibnamefont
  {Axler}},\ }\href@noop {} {\emph {\bibinfo {title} {Linear algebra done
  right}}}\ (\bibinfo  {publisher} {Springer},\ \bibinfo {year}
  {2024})\BibitemShut {NoStop}%
\bibitem [{tik(1963)}]{tikhonov1963solution}%
  \BibitemOpen
  \href@noop {} {\bibfield  {journal} {\bibinfo  {journal} {Soviet Math.
  Dokl.}\ }\textbf {\bibinfo {volume} {4}},\ \bibinfo {pages} {1035} (\bibinfo
  {year} {1963})}\BibitemShut {NoStop}%
\bibitem [{\citenamefont {Tikhonov}\ and\ \citenamefont
  {Arsenin}(1977)}]{tikhonov1977solutions}%
  \BibitemOpen
  \bibfield  {author} {\bibinfo {author} {\bibfnamefont {A.~N.}\ \bibnamefont
  {Tikhonov}}\ and\ \bibinfo {author} {\bibfnamefont {V.~Y.}\ \bibnamefont
  {Arsenin}},\ }\href@noop {} {\bibfield  {journal} {\bibinfo  {journal} {(No
  Title)}\ } (\bibinfo {year} {1977})}\BibitemShut {NoStop}%
\bibitem [{\citenamefont {Virtanen}\ \emph {et~al.}(2020)\citenamefont
  {Virtanen}, \citenamefont {Gommers}, \citenamefont {Oliphant}, \citenamefont
  {Haberland}, \citenamefont {Reddy}, \citenamefont {Cournapeau}, \citenamefont
  {Burovski}, \citenamefont {Peterson}, \citenamefont {Weckesser},
  \citenamefont {Bright} \emph {et~al.}}]{virtanen2020scipy}%
  \BibitemOpen
  \bibfield  {author} {\bibinfo {author} {\bibfnamefont {P.}~\bibnamefont
  {Virtanen}}, \bibinfo {author} {\bibfnamefont {R.}~\bibnamefont {Gommers}},
  \bibinfo {author} {\bibfnamefont {T.~E.}\ \bibnamefont {Oliphant}}, \bibinfo
  {author} {\bibfnamefont {M.}~\bibnamefont {Haberland}}, \bibinfo {author}
  {\bibfnamefont {T.}~\bibnamefont {Reddy}}, \bibinfo {author} {\bibfnamefont
  {D.}~\bibnamefont {Cournapeau}}, \bibinfo {author} {\bibfnamefont
  {E.}~\bibnamefont {Burovski}}, \bibinfo {author} {\bibfnamefont
  {P.}~\bibnamefont {Peterson}}, \bibinfo {author} {\bibfnamefont
  {W.}~\bibnamefont {Weckesser}}, \bibinfo {author} {\bibfnamefont
  {J.}~\bibnamefont {Bright}},  \emph {et~al.},\ }\href@noop {} {\bibfield
  {journal} {\bibinfo  {journal} {Nature methods}\ }\textbf {\bibinfo {volume}
  {17}},\ \bibinfo {pages} {261} (\bibinfo {year} {2020})}\BibitemShut
  {NoStop}%
\bibitem [{\citenamefont {Yang}(2010)}]{yang2010nonlinear}%
  \BibitemOpen
  \bibfield  {author} {\bibinfo {author} {\bibfnamefont {J.}~\bibnamefont
  {Yang}},\ }\href@noop {} {\emph {\bibinfo {title} {Nonlinear waves in
  integrable and nonintegrable systems}}}\ (\bibinfo  {publisher} {SIAM},\
  \bibinfo {year} {2010})\BibitemShut {NoStop}%
\bibitem [{\citenamefont {Frigo}\ and\ \citenamefont
  {Johnson}(2005)}]{frigo2005design}%
  \BibitemOpen
  \bibfield  {author} {\bibinfo {author} {\bibfnamefont {M.}~\bibnamefont
  {Frigo}}\ and\ \bibinfo {author} {\bibfnamefont {S.~G.}\ \bibnamefont
  {Johnson}},\ }\href@noop {} {\bibfield  {journal} {\bibinfo  {journal}
  {Proceedings of the IEEE}\ }\textbf {\bibinfo {volume} {93}},\ \bibinfo
  {pages} {216} (\bibinfo {year} {2005})}\BibitemShut {NoStop}%
\bibitem [{\citenamefont {Chabchoub}\ \emph
  {et~al.}(2012{\natexlab{a}})\citenamefont {Chabchoub}, \citenamefont
  {Hoffmann}, \citenamefont {Onorato},\ and\ \citenamefont
  {Akhmediev}}]{chabchoub2012super}%
  \BibitemOpen
  \bibfield  {author} {\bibinfo {author} {\bibfnamefont {A.}~\bibnamefont
  {Chabchoub}}, \bibinfo {author} {\bibfnamefont {N.}~\bibnamefont {Hoffmann}},
  \bibinfo {author} {\bibfnamefont {M.}~\bibnamefont {Onorato}}, \ and\
  \bibinfo {author} {\bibfnamefont {N.}~\bibnamefont {Akhmediev}},\ }\href@noop
  {} {\bibfield  {journal} {\bibinfo  {journal} {Physical Review X}\ }\textbf
  {\bibinfo {volume} {2}},\ \bibinfo {pages} {011015} (\bibinfo {year}
  {2012}{\natexlab{a}})}\BibitemShut {NoStop}%
\bibitem [{\citenamefont {Chabchoub}\ \emph
  {et~al.}(2012{\natexlab{b}})\citenamefont {Chabchoub}, \citenamefont
  {Hoffmann}, \citenamefont {Onorato}, \citenamefont {Slunyaev}, \citenamefont
  {Sergeeva}, \citenamefont {Pelinovsky},\ and\ \citenamefont
  {Akhmediev}}]{chabchoub2012observation}%
  \BibitemOpen
  \bibfield  {author} {\bibinfo {author} {\bibfnamefont {A.}~\bibnamefont
  {Chabchoub}}, \bibinfo {author} {\bibfnamefont {N.}~\bibnamefont {Hoffmann}},
  \bibinfo {author} {\bibfnamefont {M.}~\bibnamefont {Onorato}}, \bibinfo
  {author} {\bibfnamefont {A.}~\bibnamefont {Slunyaev}}, \bibinfo {author}
  {\bibfnamefont {A.}~\bibnamefont {Sergeeva}}, \bibinfo {author}
  {\bibfnamefont {E.}~\bibnamefont {Pelinovsky}}, \ and\ \bibinfo {author}
  {\bibfnamefont {N.}~\bibnamefont {Akhmediev}},\ }\href@noop {} {\bibfield
  {journal} {\bibinfo  {journal} {Physical Review E—Statistical, Nonlinear,
  and Soft Matter Physics}\ }\textbf {\bibinfo {volume} {86}},\ \bibinfo
  {pages} {056601} (\bibinfo {year} {2012}{\natexlab{b}})}\BibitemShut
  {NoStop}%
\bibitem [{\citenamefont {Elachi}\ and\ \citenamefont
  {Van~Zyl}(2021)}]{elachi2021introduction}%
  \BibitemOpen
  \bibfield  {author} {\bibinfo {author} {\bibfnamefont {C.}~\bibnamefont
  {Elachi}}\ and\ \bibinfo {author} {\bibfnamefont {J.~J.}\ \bibnamefont
  {Van~Zyl}},\ }\href@noop {} {\emph {\bibinfo {title} {Introduction to the
  physics and techniques of remote sensing}}}\ (\bibinfo  {publisher} {John
  Wiley \& Sons},\ \bibinfo {year} {2021})\BibitemShut {NoStop}%
\bibitem [{\citenamefont {Campagnola}\ and\ \citenamefont
  {Loew}(2003)}]{campagnola2003second}%
  \BibitemOpen
  \bibfield  {author} {\bibinfo {author} {\bibfnamefont {P.~J.}\ \bibnamefont
  {Campagnola}}\ and\ \bibinfo {author} {\bibfnamefont {L.~M.}\ \bibnamefont
  {Loew}},\ }\href@noop {} {\bibfield  {journal} {\bibinfo  {journal} {Nature
  biotechnology}\ }\textbf {\bibinfo {volume} {21}},\ \bibinfo {pages} {1356}
  (\bibinfo {year} {2003})}\BibitemShut {NoStop}%
\bibitem [{\citenamefont {Hu}\ \emph {et~al.}(2020)\citenamefont {Hu},
  \citenamefont {Field}, \citenamefont {Kelkar}, \citenamefont {Chiang},
  \citenamefont {Wernsing}, \citenamefont {Toussaint}, \citenamefont
  {Bartels},\ and\ \citenamefont {Popescu}}]{hu2020harmonic}%
  \BibitemOpen
  \bibfield  {author} {\bibinfo {author} {\bibfnamefont {C.}~\bibnamefont
  {Hu}}, \bibinfo {author} {\bibfnamefont {J.~J.}\ \bibnamefont {Field}},
  \bibinfo {author} {\bibfnamefont {V.}~\bibnamefont {Kelkar}}, \bibinfo
  {author} {\bibfnamefont {B.}~\bibnamefont {Chiang}}, \bibinfo {author}
  {\bibfnamefont {K.}~\bibnamefont {Wernsing}}, \bibinfo {author}
  {\bibfnamefont {K.~C.}\ \bibnamefont {Toussaint}}, \bibinfo {author}
  {\bibfnamefont {R.~A.}\ \bibnamefont {Bartels}}, \ and\ \bibinfo {author}
  {\bibfnamefont {G.}~\bibnamefont {Popescu}},\ }\href@noop {} {\bibfield
  {journal} {\bibinfo  {journal} {Nature Photonics}\ }\textbf {\bibinfo
  {volume} {14}},\ \bibinfo {pages} {564} (\bibinfo {year} {2020})}\BibitemShut
  {NoStop}%
\bibitem [{\citenamefont {Hong}\ and\ \citenamefont
  {Xu}(2002)}]{hong2002nondestructive}%
  \BibitemOpen
  \bibfield  {author} {\bibinfo {author} {\bibfnamefont {J.}~\bibnamefont
  {Hong}}\ and\ \bibinfo {author} {\bibfnamefont {A.}~\bibnamefont {Xu}},\
  }\href@noop {} {\bibfield  {journal} {\bibinfo  {journal} {Applied physics
  letters}\ }\textbf {\bibinfo {volume} {81}},\ \bibinfo {pages} {4868}
  (\bibinfo {year} {2002})}\BibitemShut {NoStop}%
\bibitem [{\citenamefont {Khatri}\ \emph {et~al.}(2008)\citenamefont {Khatri},
  \citenamefont {Daraio},\ and\ \citenamefont {Rizzo}}]{khatri2008highly}%
  \BibitemOpen
  \bibfield  {author} {\bibinfo {author} {\bibfnamefont {D.}~\bibnamefont
  {Khatri}}, \bibinfo {author} {\bibfnamefont {C.}~\bibnamefont {Daraio}}, \
  and\ \bibinfo {author} {\bibfnamefont {P.}~\bibnamefont {Rizzo}},\ }in\
  \href@noop {} {\emph {\bibinfo {booktitle} {Nondestructive Characterization
  for Composite Materials, Aerospace Engineering, Civil Infrastructure, and
  Homeland Security 2008}}},\ Vol.\ \bibinfo {volume} {6934}\ (\bibinfo
  {organization} {SPIE},\ \bibinfo {year} {2008})\ pp.\ \bibinfo {pages}
  {225--232}\BibitemShut {NoStop}%
\bibitem [{\citenamefont {Li}\ \emph {et~al.}(2013)\citenamefont {Li},
  \citenamefont {Yu},\ and\ \citenamefont {Yang}}]{li2013solitary}%
  \BibitemOpen
  \bibfield  {author} {\bibinfo {author} {\bibfnamefont {F.}~\bibnamefont
  {Li}}, \bibinfo {author} {\bibfnamefont {L.}~\bibnamefont {Yu}}, \ and\
  \bibinfo {author} {\bibfnamefont {J.}~\bibnamefont {Yang}},\ }\href@noop {}
  {\bibfield  {journal} {\bibinfo  {journal} {Journal of Physics D: Applied
  Physics}\ }\textbf {\bibinfo {volume} {46}},\ \bibinfo {pages} {155106}
  (\bibinfo {year} {2013})}\BibitemShut {NoStop}%
\bibitem [{\citenamefont {Jalali}\ \emph {et~al.}(2022)\citenamefont {Jalali},
  \citenamefont {Misra}, \citenamefont {Dickerson},\ and\ \citenamefont
  {Rizzo}}]{jalali2022detection}%
  \BibitemOpen
  \bibfield  {author} {\bibinfo {author} {\bibfnamefont {H.}~\bibnamefont
  {Jalali}}, \bibinfo {author} {\bibfnamefont {R.}~\bibnamefont {Misra}},
  \bibinfo {author} {\bibfnamefont {S.~J.}\ \bibnamefont {Dickerson}}, \ and\
  \bibinfo {author} {\bibfnamefont {P.}~\bibnamefont {Rizzo}},\ }\href@noop {}
  {\bibfield  {journal} {\bibinfo  {journal} {Research in Nondestructive
  Evaluation}\ }\textbf {\bibinfo {volume} {33}},\ \bibinfo {pages} {78}
  (\bibinfo {year} {2022})}\BibitemShut {NoStop}%
\bibitem [{\citenamefont {Yoon}\ \emph {et~al.}(2023)\citenamefont {Yoon},
  \citenamefont {Cantwell}, \citenamefont {Yeun}, \citenamefont {Cho},
  \citenamefont {Byon}, \citenamefont {Kim} \emph {et~al.}}]{yoon2023defect}%
  \BibitemOpen
  \bibfield  {author} {\bibinfo {author} {\bibfnamefont {S.}~\bibnamefont
  {Yoon}}, \bibinfo {author} {\bibfnamefont {W.~J.}\ \bibnamefont {Cantwell}},
  \bibinfo {author} {\bibfnamefont {C.~Y.}\ \bibnamefont {Yeun}}, \bibinfo
  {author} {\bibfnamefont {C.-S.}\ \bibnamefont {Cho}}, \bibinfo {author}
  {\bibfnamefont {Y.-J.}\ \bibnamefont {Byon}}, \bibinfo {author}
  {\bibfnamefont {T.-Y.}\ \bibnamefont {Kim}},  \emph {et~al.},\ }\href@noop {}
  {\bibfield  {journal} {\bibinfo  {journal} {International Journal of
  Mechanical Sciences}\ }\textbf {\bibinfo {volume} {239}},\ \bibinfo {pages}
  {107882} (\bibinfo {year} {2023})}\BibitemShut {NoStop}%
\end{thebibliography}

%merlin.mbs apsrev4-1.bst 2010-07-25 4.21a (PWD, AO, DPC) hacked
%Control: key (0)
%Control: author (72) initials jnrlst
%Control: editor formatted (1) identically to author
%Control: production of article title (-1) disabled
%Control: page (0) single
%Control: year (1) truncated
%Control: production of eprint (0) enabled
%

\end{document}